\documentclass[final, 1p, times, review, 10pt]{elsarticle}

\usepackage{amssymb}
\usepackage{graphicx,subfigure}
\usepackage{amsmath}
\usepackage{amsfonts}
\usepackage{enumerate}
\usepackage{amsthm}
\usepackage{epsfig}
\usepackage{bm}
\usepackage{booktabs}
\usepackage{epstopdf} % 将图片从eps格式转换为pdf格式

\usepackage{lineno}
\usepackage{caption}
\usepackage{setspace}

\biboptions{sort&compress}

\begin{document}
\begin{frontmatter}
\title{Coherent Integration for Targets with Constant Cartesian Velocities Based on Accurate Range Model}

\author{Gongjian Zhou, Zeyu Xu, and Yuchao Yang}

\address{School of Electronics and Information Engineering, Harbin Institute of Technology, Harbin, China \\
Key Laboratory of Marine Environmental Monitoring and Information Processing, Ministry of Industry and Information Technology, China \\
Corresponding author: Gongjian Zhou \\
Email: zhougj@hit.edu.cn 
}

\begin{abstract}
Long-time coherent integration (LTCI) is one of the most important techniques to improve radar detection performance of weak targets. However, for the targets moving with constant Cartesian velocities (CCV), the existing LTCI methods based on polynomial motion models suffer from limited integration time and coverage of target speed due to model mismatch. Here, a novel generalized Radon Fourier transform method for CCV targets is presented, based on the accurate range evolving model, which is a square root of a polynomial with terms up to the second order with target speed as the factor. The accurate model instead of approximate polynomial models used in the proposed method enables effective energy integration on characteristic invariant with feasible computational complexity. The target samplings are collected and the phase fluctuation among pulses is compensated according to the accurate range model. The high order range migration and complex Doppler frequency migration caused by the highly nonlinear signal are eliminated simultaneously. Integration results demonstrate that the proposed method can not only achieve effective coherent integration of CCV targets regardless of target speed and coherent processing interval, but also provide additional observation and resolution in speed domain. 
\end{abstract}


\end{frontmatter}

% \linenumbers

\section{Introduction}\label{Introduction}
To improve the detection performance of targets 
in a large coverage of position and target characteristic via pulse integration is a hot topic 
in the area of radar technology \cite{multi-target-detection-2010, KTME-2011, RLCT-2014, SDFC-LVT-2016, RFPPT-2017, TRT-SGRFT-2017, SAF-SFT-2019, BRFC-RPSCT-2019, PSPLVD-2020, Filterbank-Framework-2020}. 
It is known that coherent integration can produce better performance than 
the noncoherent integration by compensating phase fluctuation among sampling pulses \cite{CI1-2008, CI2-2003, RFAF-2017}. 
The moving target detection (MTD) method \cite{MTD1-2002, MTD2-2001} has been widely used in modern radar systems to 
suppress strong background clutter and integrate energy of target echo signal with unknown Doppler frequency. 
However, the performance gain of the MTD is limited by the target's resident in a range cell. 
The range migration (RM) and Doppler frequency migration (DFM) effects may occur when target performs maneuvers, 
e.g., high velocity and acceleration, during the integration interval. 
This prevents the MTD method from improving detection performance through long-time coherent integration (LTCI).

In order to realize effective detection of 
low signal-to-noise ratio (SNR) targets (e.g., far-range or stealth targets) 
and high maneuvering targets (e.g., ballistic missiles, jet fighters), 
LTCI methods with ability of RM correction and/or DFM compensation have been under intensive investigation for the past decade. 
The related works can be classified into three categories based on the target motion model considered in the methods.

In the first category, the target motion model is formulated by a primary polynomial with respect to slow time. 
In this case, the effect of RM should be eliminated to obtain effective integration for a quite long time. 
The most popular methods may be keystone transform (KT) \cite{KT1-1999, KT2-2008, KT3-2007} 
and Radon Fourier transform (RFT) \cite{RFT1-2011, RFT2-2011, RFT3-2012}. 
The KT method corrects the RM effect by rescaling the time axis for each frequency, 
while the RFT method realizes coherent integration for moving targets with RM effect by 
a two-dimensional searching scheme in range-Doppler domain. 
Some other typical methods include sequence reversing transform (SRT) \cite{SRT-2017}, 
axis rotation MTD (AR-MTD) \cite{AR-MTD-2014}, 
scaled inverse Fourier transform (SCIFT) \cite{SCIFT-2015}, 
frequency-domain deramp-KT (FDDKT) \cite{FDDKT-2016}, 
modified location rotation transform (MLRT) \cite{MLRT-2018}, etc. 
In these methods \cite{KT1-1999, KT2-2008, KT3-2007, RFT1-2011, RFT2-2011, RFT3-2012, SRT-2017, AR-MTD-2014, SCIFT-2015, FDDKT-2016, MLRT-2018}, 
the target is assumed to move with a constant radial velocity during the integration interval, 
which is actually a very strict assumption and will be violated easily in practical applications.

In the second category, the target motion model is described by a quadratic polynomial, 
where the target is assumed to move with a constant radial acceleration. 
In this case, not only the RM caused by target's radial velocity, 
but also the DFM and range curvature (RC) induced by target's radial acceleration would occur. 
The improved axis rotation-fractional Fourier transform (IAR-FRFT) \cite{IAR-FRFT-2015} 
and KT with Lv's distribution (KT-LVD) \cite{KT-LVD-2016} 
were proposed to eliminate the linear RM and DFM effects. 
To further improve the detection performance of target with constant radial acceleration, 
some more effective methods have been proposed, 
including radon-fractional Fourier transform (RFRFT) \cite{RFRFT-2014}, 
KT and matched filtering process (KT-MFP) \cite{KT-MFP-2017}, 
Radon-Lv's distribution (RLVD) \cite{RLVD-2015}, 
two steps scaling and fractional Fourier transform (TSS-FRFT) \cite{TSS-FrFT-2019}, etc. 
In these methods \cite{RFRFT-2014, KT-MFP-2017, RLVD-2015, TSS-FrFT-2019}, 
the RM induced by both the target's radial velocity and radial acceleration are corrected.

In the third category, more complex motion is considered by formulating the target motion with constant radial jerk, 
where the target motion model is expressed by a cubic polynomial. 
RM effects up to third order and DFM effects up to second order can be induced by this motion. 
In dealing with LTCI problem in this case, 
the generalized RFT (GRFT) \cite{GRFT-2012} 
and polynomial Radon-polynomial Fourier transform (PRPFT) \cite{PRPFT-2018} 
could obtain optimal integration gain, 
but may be computationally prohibitive due to the high dimensional searching strategy. 
The adjacent cross correlation function (ACCF) \cite{ACCF1-2015, ACCF2-2015, ACCF3-2016} 
is more computationally effective at a cost of performance degradation. 
Some other methods, such as generalized KT and generalized dechirp process (GKTGDP) \cite{GKTGDP-2015}, 
KT and generalized dechirp process (KTGDP) \cite{KTGDP-2016}, 
KT and matched filtering processing \cite{KTMF-2016}, 
Radon modified Lv's distribution (RMLVD) \cite{RMLVD-2017}, 
Radon-high-order time-chirp rate transform (RHTRT) \cite{RHTRT-2016}, 
discrete polynomial-phase transform and Lv's distribution \cite{DPT-LVD-2018}, 
the method based on series reversion \cite{SR-2019} were also proposed, 
where the tradeoff between integration performance and computational complexity is considered.

In the aforementioned methods \cite{KT1-1999, KT2-2008, KT3-2007, RFT1-2011, RFT2-2011, RFT3-2012, SRT-2017, AR-MTD-2014, SCIFT-2015, FDDKT-2016, MLRT-2018, IAR-FRFT-2015, KT-LVD-2016, RFRFT-2014, KT-MFP-2017, RLVD-2015, TSS-FrFT-2019, GRFT-2012, PRPFT-2018, ACCF1-2015, ACCF2-2015, ACCF3-2016, GKTGDP-2015, KTGDP-2016, KTMF-2016, RMLVD-2017, RHTRT-2016, DPT-LVD-2018, SR-2019}, 
the target motion model is formulated straightly in range coordinate, 
based on finite order polynomial with respect to slow time. 
However, the time evolving function of a real target moving in the Cartesian coordinates is highly nonlinear, 
for example the motion with constant Cartesian velocities (CCV), 
due to the nonlinearity between Cartesian and polar coordinates. 
The polynomial motion models mentioned above may not be able to 
provide accurate description of a target with the common CCV motion. 
Actually, in some literatures \cite{PRPFT-2018, RMLVD-2017, RHTRT-2016, DPT-LVD-2018}, 
approximations based on Taylor series expansion are explicitly employed, 
resulting in the polynomial models. 
To achieve accurate formulation, 
the polynomial model with infinite orders is desired, 
but this is impractical. 
The existing LTCI methods based on finite order polynomial models cannot guarantee optimal integration performance 
due to model mismatch and inaccurate energy accumulation.

In this paper, a novel generalized Radon Fourier transform method is presented by 
formulating the signal model directly by the accurate range evolving equation corresponding to the CCV motion, 
without approximations based on Taylor series expansion. 
The target samplings are extracted according to the accurate range evolving model 
and the phase fluctuation among different pulses is compensated, 
which eliminates the high-order RM and complex DFM simultaneously. 
Since there is no error in the range model 
and the factor of highest order in the model is related to the invariant speed of the CCV target, 
the proposed GRFT based on accurate range evolving model (AREM-GRFT) can achieve effective coherent integration 
regardless of target speed and coherent processing interval (CPI). 
Moreover, the AREM-GRFT method can provide additional observation and resolution of the target in speed domain by 
focusing on the motion characteristic invariant, 
which is of significance for both target detection and tracking. 
Simulation results demonstrate the effectiveness and superiority 
of the proposed AREM-GRFT method 
in terms of integration performance and detection ability.

The remainder of this paper is organized as follows. 
In Section \ref{Problem Formulation}, the signal model is established. 
The AREM-GRFT method is presented in Section \ref{Long Time Coherent Integration via AREM-GRFT}. 
In Section \ref{Numerical Simulation Experiments}, some numerical simulation experiments are provided to 
evaluate the performance of the proposed method, followed by conclusions in Section \ref{Conclusion}.

%%%%%%%%%%%%%%%%%%%%%%%%%%%%%%%%%%%%%%%%%%%%%%%%%%%%%%%%%%%%%%%%%%%%%%%%%%%%%%%%%%%%%%
\section{Problem formulation}\label{Problem Formulation}
Suppose that the radar transmits the linear frequency modulated (LFM) signal, i.e., 
\begin{equation}
{s_T} \left( {\tilde t} \right) = {\rm{rect}} \left( { \frac{\tilde t}{T_p} } \right) \exp \left( {j\pi \mu {{\tilde t}^2}} \right) \exp \left( {j2\pi {f_c} \tilde t} \right)
\end{equation}
where $\operatorname{rect}(u) = \Bigg\{\begin{array}{ll}{1,} & {|u| \leq 1/2 } \vspace{-1.5ex} \\ {0,} & {|u| > 1/2}\end{array}\Bigg.$, 
$\tilde t$ denotes the fast time, 
${f_c}$ denotes the carrier frequency, 
${T_p}$ denotes the pulse duration, 
$\mu=B / T_{p}$ represents the frequency modulation rate, 
and $B$ denotes the signal bandwidth.

The received baseband echo signal of a moving target can be expressed as \cite{RFT1-2011, GRFT-2012} 
\begin{equation}
{s_r} \left( { {t_m}, \tilde{t} } \right) = {A_1}{\rm{rect}} \left( { \dfrac{\tilde t - \tau }{T_p} } \right) \exp \left[ { j\pi \mu {{\left( {\tilde t - \tau } \right)}^2} } \right] \exp \left( { - j2 \pi {f_c} \tau } \right)
\end{equation}
where ${A_1}$ is target reflectivity, 
$\tau=2 r\left(t_{m}\right) / c$ is the time delay, 
$c$ is the speed of light, 
$r\left( {t_m} \right)$ is the instantaneous slant range between the moving target and the radar, 
${t_m} = m{T_r}\left( {m = 0,1, \ldots ,M - 1} \right)$ denotes the slow time, 
$M$ and ${T_r}$ represent the number of integration pulses and pulse repetition time, respectively.

After pulse compression (PC), the compressed signal of the moving target can be written as \cite{RFT1-2011, ACCF3-2016}
\begin{equation} \label{original PC signal}
s \left( { {t_m}, \tilde{t} } \right) = {A_2}{\rm{sinc}} \left[ {B \left( {\tilde t - \dfrac{ 2r \left( {t_m} \right) }{c} } \right)} \right] \exp \left( - j \frac{4 \pi}{\lambda} r \left( t_m \right) \right)
\end{equation}
where $\operatorname{sinc}(x) = \sin (\pi x) /(\pi x)$ denotes the sinc function, 
${A_2}$ denotes the signal amplitude after PC, 
and $\lambda=c / f_{c}$ denotes the wavelength.

Let $\tilde{t} = 2 \tilde{r} / c$, (\ref{original PC signal}) can be further written as \cite{RFT1-2011}
\begin{equation} \label{PC_rewritten}
s \left( { {t_m}, \tilde{r} } \right) = {A_2}{\rm{sinc}} \left[ {B \left( \dfrac{ 2 \left( \tilde{r} - r \left( t_m \right) \right) }{c} \right) } \right] \exp \left( - j \frac{4 \pi}{\lambda} r \left( t_m \right) \right)
\end{equation}
where $\tilde{r}$ is the range corresponding to the fast time $\tilde{t}$.

It is seen from (\ref{PC_rewritten}) that 
the envelope position and phase of the target compressed signal are totally determined by the target range. 
Due to the target movement, 
the signal envelope would be shifted away from its original position with the increase of the integration time. 
When the offset of the signal envelope exceeds the range resolution $\rho_{r}=c /(2 B)$, 
the RM effect would occur \cite{RFRAF-2015}. 
In addition, the change of range rate will also cause the offset of Doppler spectrum 
(the instantaneous Doppler frequency can be expressed as $f_{d} (t_m) = -\frac{2}{\lambda} \frac{\text{d} r(t_m)}{\text{d} t_m}$). 
When the offset of the Doppler spectrum exceeds the Doppler resolution $\rho_{d}=1 /(M T_{r})$, 
the DFM effect would happen \cite{RFRFT-2014}. 
The migration effects would make it difficult to coherently integrate target's signal energy, 
which may deteriorate target detection performance. 
In order to eliminate the RM and DFM effects due to target movement, 
it is crucial to establish an accurate range evolution model, 
which needs to match the actual movement of the target. 
As mentioned in the previous section, 
the existing LTCI methods differ in the order of the polynomial model of $r\left( {t_m} \right)$. 
Nevertheless, the first-order, the second-order polynomial models 
and even the higher order polynomial models may not match the CCV motion well, 
which makes it difficult for LTCI methods based on these polynomial models to 
achieve effective coherent integration of CCV targets. 
Therefore, the accurate range evolving model instead of approximate polynomial model is 
used in this paper to formulate the echo signal, 
corresponding to the target moving with constant Cartesian velocities.

\begin{figure}[!t]  
	\centering  
	\includegraphics[scale= 0.85 ]{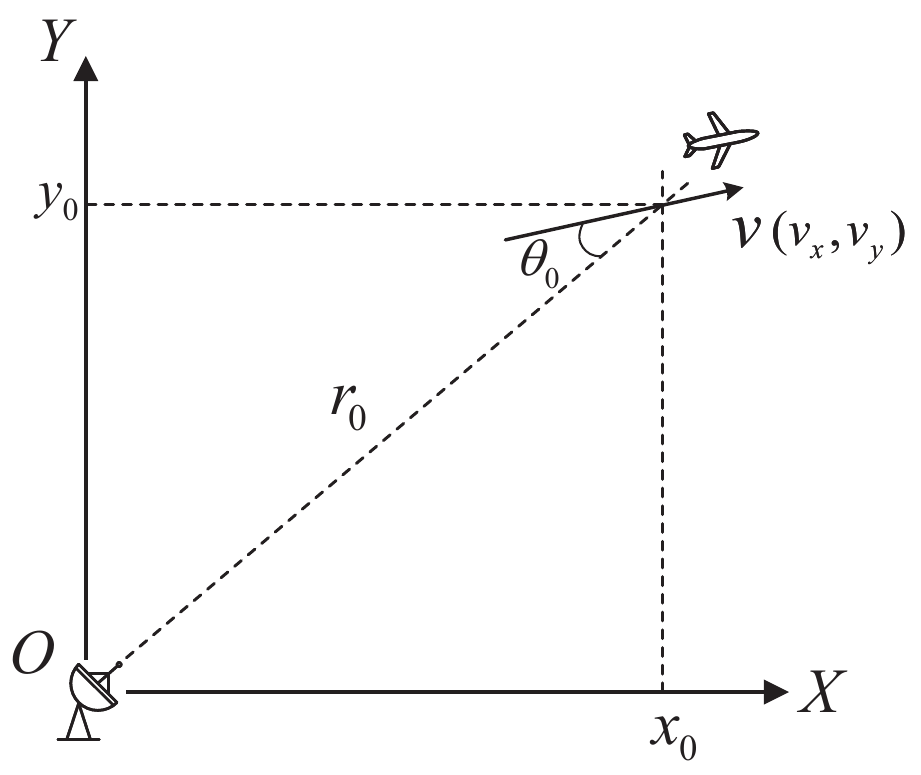}  
	\caption{Target motion with constant Cartesian velocities.}   
	\label{CCV target motion model}
\end{figure}

%%%%%%%%%%%%%%%%%%%%%%%%%%%%%%%%%%%%%%%%%%%%%%%%%%%%%%%%%%%%%%%%%%%%%%%%%%%%%%%%%%%%%%
\section{Long time coherent integration via AREM-GRFT}\label{Long Time Coherent Integration via AREM-GRFT}

\subsection{Accurate range evolving model}\label{Accurate Range Evolving Model}
As shown in Fig. \ref{CCV target motion model}, 
suppose that there is a target moving with constant Cartesian velocities $\left( {{v_x}, {v_y}} \right)$, 
starting from position $\left( {{x_0}, {y_0}} \right)$ in Cartesian coordinates. 
A radar is observing the target at the origin of the coordinates.

According to the assumption of constant velocity motion in Cartesian coordinates, 
the instantaneous Cartesian position of the target in x and y directions can be expressed as 
\begin{equation} \label{instantaneous Cartesian position}
\left\{
\begin{array}{lr} 
{x \left( {t_m} \right) = {x_0} + {v_x}{t_m}} &  \vspace{-1.5ex} \\ 
{y \left( {t_m} \right) = {y_0} + {v_y}{t_m}} &  
\end{array}\right.
\end{equation}
Then, the instantaneous slant range $r\left( {t_m} \right)$ between the target and radar 
during the observation time (i.e., $0 \le {t_m} \le \rm{CPI}$) can be given by 
\begin{equation} \label{instantaneous slant range}
\begin{split}
r \left( t_m \right) =& \sqrt { \left[ x{\left( {t_m} \right)} \right]^2 + \left[ y{\left( {t_m} \right)} \right]^2} \\
=& \sqrt { x_0^2 + y_0^2 + 2\left( {{x_0}{v_x} + {y_0}{v_y}} \right){t_m} + \left( {v_x^2 + v_y^2} \right)t_m^2}
\end{split}
\end{equation}

The initial slant range between the target and radar is 
\begin{equation} \label{initial slant range}
{r_0} = \sqrt {x_0^2 + y_0^2}
\end{equation}

The instantaneous radial velocity (also called range rate or Doppler velocity) of the target relative to the radar can be expressed as 
\begin{equation}
\dot{r}\left(t_{m}\right) = \dfrac{ \text{d} r(t_{m})}{ \text{d} t_{m}} = \dfrac{x\left(t_{m}\right) \dot{x}\left(t_{m}\right) + y\left(t_{m}\right) \dot{y}\left(t_{m}\right)}{r\left(t_{m}\right)}
\end{equation}
where $\dot{x}\left(t_{m}\right)$ and $\dot{y}\left(t_{m}\right)$ denote 
the instantaneous Cartesian velocities of the target in x and y directions, respectively. 
For CCV motion, the Cartesian velocities are time-invariant. 
Therefore, we denote $\dot{x}\left(t_{m}\right)$ and $\dot{y}\left(t_{m}\right)$ by $v_x$ and $v_y$, respectively.

Similarly, the initial radial velocity of the target relative to the radar is 
\begin{equation} \label{initial radial velocity}
{\dot r_0} = \dfrac{{x_0}{v_x} + {y_0}{v_y}}{\sqrt {x_0^2 + y_0^2} }
\end{equation}

Substituting (\ref{initial slant range}) and (\ref{initial radial velocity}) into (\ref{instantaneous slant range}), 
the time evolution of target range can be expressed as 
\begin{equation} \label{accurate range evolving model}
r\left( {t_m} \right) = \sqrt {r_0^2 + 2{r_0}{{\dot r}_0}{t_m} + {v^2}t_m^2}
\end{equation}
where $v = \sqrt {v_x^2 + v_y^2}$ denotes the invariant speed of the CCV target.

It can be seen from (\ref{accurate range evolving model}) that 
the time evolution of range for a CCV target is a highly nonlinear function 
with the square root over the second order polynomial. 
Note that although the accurate range evolving equation in (\ref{accurate range evolving model}) is derived in a two-dimensional Cartesian coordinates, 
the same equation can be obtained in the three-dimensional Cartesian coordinates. 
The polynomial motion models, 
such as the first-order polynomial model, 
the second-order polynomial model, 
fail to accurately describe the model in (\ref{accurate range evolving model}). 
Actually, in order to represent the model in (\ref{accurate range evolving model}) by polynomial, 
Taylor series expansion with infinite terms is required as 
\begin{equation} \label{Taylor series expansion with infinite terms}
\begin{split}
r\left( {t_m} \right) =& \sum_{l=0}^{\infty} \frac{r^{(l)}(0)}{l !} t_m^l  \\
=& {r_0} + {\dot r_0}{t_m} + \left( \dfrac{{v^2} - \dot r_0^2}{2{r_0}} \right)t_m^2 + \left( \dfrac{\dot r_0^3 - {{\dot r}_0}{v^2}}{2r_0^2} \right)t_m^3 + \left( \dfrac{6\dot r_0^2{v^2} - {v^4} - 5\dot r_0^4}{8r_0^3} \right)t_m^4 +  \cdots 
\end{split}
\end{equation}
where $r^{(l)}(0)$, $l$ $=$ $0$, $1$, $2$, $\ldots$ , denotes the $l$th-order derivative of $r\left( {t_m} \right)$ at $t_m=0$.

The high nonlinearity of $r\left( {t_m} \right)$, 
as illustrated in (\ref{accurate range evolving model}) and (\ref{Taylor series expansion with infinite terms}), 
will cause high-order RM and complex DFM effects, 
which may not be handled well by the LTCI methods based on polynomial model with finite terms. 
The discarding of any high order in (\ref{Taylor series expansion with infinite terms}) may lead to model mismatch and performance degradation, 
especially in the case of high speed and extremely long integration time. 
In addition, since the target range and radial velocity are time variant, 
the factor of each term in (\ref{Taylor series expansion with infinite terms}) differs in different integration intervals, 
which is not good for data association and target tracking. 
The factor of the second-order term in (\ref{Taylor series expansion with infinite terms}) is not a real range acceleration. 
In other words, for a CCV target, 
the second-order polynomial based LTCI methods \cite{IAR-FRFT-2015, RFRFT-2014, KT-MFP-2017} may not 
provide a parameter estimation with a physical meaning of range acceleration. 
On the contrary, the factor of the second order of the polynomial under the square root in (\ref{accurate range evolving model}) has 
a physical meaning of speed square and is time invariant for CCV targets.

In this paper, 
the accurate time evolution of range in (\ref{accurate range evolving model}) is directly used to 
formulate the signal model without approximations to eliminate the problem of model mismatch. 
And a GRFT based method is presented to realize effective coherent integration for CCV targets 
and provide range-Doppler measurement as well as speed measurement, 
which comes from focusing on the time invariant speed.

\subsection{Definition of AREM-GRFT}\label{AREM-GRFT Algorithm}
Substituting (\ref{accurate range evolving model}) into (\ref{PC_rewritten}), 
the compressed signal of a target with CCV motion can be represented as 
\begin{equation} \label{compressed signal of a target with CCV motion}
s \left( {{t_m},\tilde r} \right) = {A_2}{\rm{sinc}} \left[ { B \left( { \dfrac{2\left( {\tilde r - \sqrt {r_0^2 + 2{r_0}{{\dot r}_0}{t_m} + {v^2}t_m^2} } \right)}{c} } \right) } \right] \exp \left( { - j \dfrac{4\pi }{\lambda} \sqrt {r_0^2 + 2{r_0}{{\dot r}_0}{t_m} + {v^2}t_m^2} } \right)
\end{equation}

As shown in (\ref{compressed signal of a target with CCV motion}), 
the envelope position and target signal phase change complicatedly with time 
due to the comprehensive effect of target's initial range, initial radial velocity and speed, 
which will cause high-order RM and complex DFM within the integration time. 
To achieve effective energy accumulation under this complex conditions, 
the AREM-GRFT method is presented, 
which is defined as 
\begin{equation} \label{the AREM-GRFT method}
G \left( {r_{s}, \dot{r}_{s}, v_{s}} \right) = \int_0^T { s\left( { t_m, \sqrt{r_{s}^{2} + 2 r_{s} \dot{r}_{s} t_m + v_{s}^{2} t_m^2} } \right) } \exp \left( { j \dfrac{ 4\pi }{\lambda} \sqrt{r_{s}^{2} + 2 r_{s} \dot{r}_{s} t_m + v_{s}^{2} t_m^2} } \right) dt_m 
\end{equation}
where $r_{s}$, $\dot{r}_{s}$ and $v_{s}$ denote respectively the searching slant range, 
searching radial velocity and searching speed, $T$ denotes the coherent integration time. 
The AREM-GRFT can be interpreted as employing the accurate range evolving model to 
extract the target signal and compensate for the phase fluctuation. 
Since the accurate range evolution model is used, 
the complex RM and DFM effects can be eliminated and effective energy accumulation can be achieved.

As is known, the definition of RFT is \cite{RFT1-2011}  
\begin{equation} \label{the definition of RFT}
RFT\left( {r_{s}, \dot{r}_{s}} \right) = \int_0^T { s\left( {t_m, r_{s} + \dot{r}_{s}t_m} \right) } \exp \left( { j \dfrac{4\pi}{\lambda} \dot{r}_{s} t_m } \right) dt_m 
\end{equation}
In the above, 
the assumption that the target moves with a constant radial velocity in range coordinate is used. 
However, this cannot be satisfied for most CCV targets, 
since the target's radial velocity usually varies and is different from target's speed. 
Actually, the RFT method is a special case of the AREM-GRFT method. 
When the target's speed equals its radial velocity, that is $\dot{r}_{0}=v$, 
the AREM-GRFT is reduced to RFT. 
The proposed AREM-GRFT method is valid for all CCV targets, 
while the assumption of the RFT is very restrict. 
In addition, when encountering multiple CCV targets with the same initial slant range and radial velocity, 
the RFT method cannot distinguish among these targets 
due to insufficient reflection of the motion characteristic for CCV target. 
Because of the introduction of the extended parameter (i.e., $v_{s}$) in AREM-GRFT method, 
these moving targets can be distinguished in the speed domain, 
as will be illustrated in the numerical simulation experiments.

For convenience, the searching trajectory based on the accurate range model is denoted as $r_{s}(t_m)$, i.e., 
\begin{equation} \label{searching trajectory based on the accurate range model}
r_{s}(t_m) = \sqrt{r_{s}^{2} + 2 r_{s} \dot{r}_{s} t_m + v_{s}^{2} t_m^2}
\end{equation}

Substituting (\ref{PC_rewritten}) and (\ref{searching trajectory based on the accurate range model}) into (\ref{the AREM-GRFT method}), 
the coherent integration output via AREM-GRFT can be expressed as 
\begin{equation} \label{proposed method with definite signal model}
G \left( {r_{s}, \dot{r}_{s}, v_{s}} \right) = \int_{0}^{T} A_{2} \operatorname{sinc} \left[ B \left( \frac{2 \left( r_{s}(t_m) - r(t_m) \right)}{c} \right) \right] \exp \left( j \dfrac{4 \pi}{\lambda} \left( r_{s}(t_m) - r(t_m) \right) \right) dt_m 
\end{equation}

As illustrated in (\ref{proposed method with definite signal model}), 
the coherent integration output via AREM-GRFT with respect to 
different searching motion parameter pairs $( {r_{s}, \dot{r}_{s}, v_{s}} )$ can be obtained. 
The effective energy integration is based on the fact that 
the RM elimination and DFM compensation are accomplished only when the searching parameters are consistent with 
the true motion parameters of the target.

When $r_{s}=r_0$, $\dot{r}_{s}={\dot{r}_0}$ and $v_{s}=v$, we have 
\begin{equation} \label{equiv zero}
\begin{split}
r_{s}(t_m) - r(t_m) &= \sqrt{r_{s}^{2} + 2 r_{s} \dot{r}_{s} t_m + v_{s}^{2} t_m^2} - \sqrt{r_0^2 + 2{r_0}{{\dot r}_0}{t_m} + {v^2}t_m^2} \\
&\equiv 0   %\quad \quad \quad \quad \quad    %(\mathrm{when} \,\, r_{s}=r_0, \dot{r}_{s}=\dot{r}, v_{s}=v)
\end{split}
\end{equation}

It can be seen from (\ref{equiv zero}) that 
$r_{s}(t_m) - r(t_m)$ does not change with slow time and is constantly equal to zero. 
Substituting (\ref{equiv zero}) into (\ref{proposed method with definite signal model}), 
the corresponding coherent integration output via AREM-GRFT can be rewritten as 
\begin{equation} \label{total output maximum}
\begin{split}
G \left( {r_{s}, \dot{r}_{s}, v_{s}} \right) &= \int_{0}^{T} A_{2} \operatorname{sinc} \left[ \frac{2 B}{c} 0 \right] \exp \left( j \dfrac{4 \pi}{\lambda} 0 \right) dt_m \\
&= A_{2}T 
\end{split}
\end{equation}

As shown in (\ref{total output maximum}), 
the high-order RM and complex DFM are eliminated simultaneously 
when the searching motion parameters match the target's motion parameters 
(i.e., $r_{s}=r_0$, $\dot{r}_{s}={\dot{r}_0}$ and $v_{s}=v$). 
As a result, the signal energy distributed in multiple range and Doppler cells is totally extracted and coherently integrated.

\subsection{Procedure of long-time coherent integration via AREM-GRFT}\label{Procedure of LTCI via AREM-GRFT}
\begin{figure}[htbp]  
	\centering  
	\includegraphics[scale= 0.47 ]{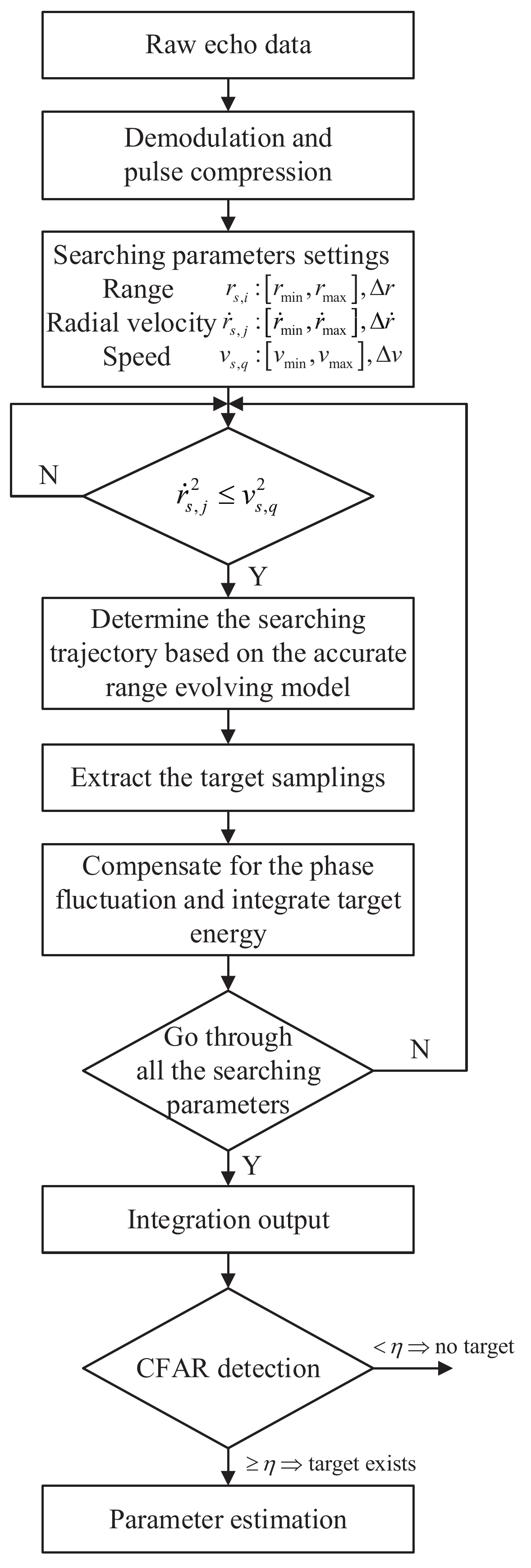}  
	\caption{Flowchart of long-time coherent integration via AREM-GRFT method.}   
	\label{Flowchart of LTCI via AREM-GRFT method}
\end{figure}

The target samplings in the range and slow time plane are extracted 
according to the accurate range evolving model with preset searching parameters. 
Then the phase fluctuation among different pulses is simultaneously compensated by 
the determined searching trajectory. 
When the searching trajectory determined by target's initial range, initial radial velocity and speed 
is the same as the true trajectory of the target, 
the signal energy of target would be focused in the range-Doppler-speed domain, 
which can eliminate the complex RM and DFM effectively. 
The flowchart of long-time coherent integration via AREM-GRFT is given in Fig. \ref{Flowchart of LTCI via AREM-GRFT method}, 
and the main steps are illustrated as follows.

\emph{Step 1: Perform demodulation and pulse compression on the raw echo data. }

During the observation time, 
the raw echo data are sampled and stored as a two-dimensional matrix in the range and slow time domain. 
After demodulation and pulse compression, 
the compressed echo data $s\left( {m,n} \right)$ for coherent integration are then obtained, 
where $m$ and $n$ denote the pulse index and discrete sampling range cell index, respectively.

\emph{Step 2: Set coherent integration parameters for AREM-GRFT. }

Based on the prior information and the motion characteristics of the target, 
the searching scopes of range, radial velocity and speed can be determined as 
$\left[ { {r_{\min }}, {r_{\max }} } \right]$, 
$\left[ {{{\dot r}_{\min }}, {{\dot r}_{\max }}} \right]$ 
and $\left[ {{v_{\min }}, {v_{\max }}} \right]$, respectively. 
According to the determined radar system parameters, 
the searching intervals of the range, radial velocity and speed can be respectively set as \cite{RFRFT-2014} 
\begin{equation}
\Delta r=c /(2 B) 
\end{equation}
\begin{equation}
\Delta \dot{r} =\lambda /(2 T) 
\end{equation}
\begin{equation}
\Delta v =\lambda /(2 T) 
\end{equation}

Accordingly, 
the discrete values of the searching range, searching radial velocity and searching speed can be respectively represented as 
\begin{equation}
{r_{s,i}} = {r_{\min }}              \,:\,   \Delta r        \,:\,    {r_{\max }},         \quad     i = 1,2, \ldots ,{N_r}
\end{equation}
\begin{equation}
{\dot r_{s,j}} = {\dot r_{\min }}    \,:\,   \Delta \dot r   \,:\,    {\dot r_{\max }},    \quad     j = 1,2, \ldots ,{N_{\dot r}}
\end{equation}
\begin{equation}
{v_{s,q}} = {v_{\min }}              \,:\,   \Delta v        \,:\,    {v_{\max }},         \quad     q = 1,2, \ldots ,{N_{v}}
\end{equation}
where $N_{r}$, $N_{\dot{r}}$ and $N_{v}$ denote the searching number of the range, radial velocity and speed, respectively.

\emph{Step 3: Apply the AREM-GRFT operation on the compressed data. }

Determine the searching trajectory according to the preset searching parameters. 
Each time a set of searching parameter pair $\left( {r_{s,i}}, {\dot r_{s,j}}, {v_{s,q}} \right)$ is determined, 
a searching trajectory based on the accurate range evolving model is determined, 
which can be expressed as 
\begin{equation} \label{discretize}
r_{s} \left( m T_r \right) = \sqrt{r_{s, i}^{2} + 2{r_{s,i}}{\dot r_{s,j}} m T_{r} + v_{s, q}^{2} m^{2} T_{r}^{2} }
\end{equation}

Extract the target samplings in the range and slow time plane according to the determined searching trajectory. 
Only when the searching motion parameters are identical to the true motion parameters of the target, 
the target signal along the actual moving trajectory can be extracted and thus the highly nonlinear RM is eliminated. 
The extracted samplings ${X_M}\left( m T_r \right)$ can be expressed as 
\begin{equation}
\begin{split}
{X_M} \left( m T_r \right) &= s \left( {m, \operatorname{round} \left( { \dfrac{r_{s} \left( m T_r \right)}{\Delta_{r}} } \right)} \right) \\
&= s \left(m, \operatorname{round} \left( \dfrac{\sqrt{r_{s, i}^{2} + 2{r_{s,i}}{\dot r_{s,j}} m T_{r} + v_{s, q}^{2} m^{2} T_{r}^{2} }}{\Delta_{r}} \right)\right)
\end{split}
\end{equation}
where $\operatorname{round}\left(  \cdot  \right)$ represents the integer operator, 
$\Delta_{r}=c / \left( 2 f_{s} \right)$ is the range sampling cell, 
$f_{s}$ is the sampling frequency.

The extracted samplings ${X_M}\left( m T_r \right)$ are compensated and summed to 
eliminate the complex DFM and integrate target energy. 
Each set of searching parameter pair $\left( {r_{s,i}}, {\dot r_{s,j}}, {v_{s,q}} \right)$ 
has a corresponding integration output $G\left( {r_{s,i}}, {\dot r_{s,j}}, {v_{s,q}} \right)$, 
which can be expressed as 
\begin{equation} \label{each set corresponding integration output}
\begin{split}
G \left( {r_{s,i}}, {\dot r_{s,j}}, {v_{s,q}} \right) &=\! \sum\limits_{m = 0}^{M - 1} {{X_M} \left( m T_r \right)} \exp \left( j \frac{4\pi}{\lambda} r_{s} \left( m T_r \right)  \right) \\
&=\! \sum\limits_{m = 0}^{M - 1} \left \{\begin{array}{ll} 
& \!\!\!\!\!\!\!\! s \left( m, \operatorname{round} \left( \dfrac{ \sqrt{r_{s, i}^{2} + 2{r_{s,i}}{\dot r_{s,j}} m T_{r} + v_{s, q}^{2} m^{2} T_{r}^{2} } }{\Delta_{r}} \right) \right) \!\! \\ 
& \!\!\!\! \exp \left( {j \dfrac{4\pi}{\lambda} \sqrt{r_{s, i}^{2} + 2{r_{s,i}}{\dot r_{s,j}} m T_{r} + v_{s, q}^{2} m^{2} T_{r}^{2} } } \right) 
\end{array} \right\}
\end{split}
\end{equation}

It can be seen from (\ref{each set corresponding integration output}) that 
the ideal integration gain of the discrete AREM-GRFT method is the number of integration pulses.

\emph{Step 4: Go through all searching parameters to obtain the integration output in range-Doppler-speed domain. }

Repeat step 3 for all the searching range, radial velocity and speed to 
obtain the output matrix in the three-dimensional domain. 
When a target exists, a peak will appear at the position with three parameters close to 
the target's initial range, initial radial velocity and speed, respectively.

\emph{Remark 1:} Since the absolute value of the target radial velocity is 
less than or equal to the absolute value of the velocity, 
the case that the searching speed is less than the absolute value of searching radial velocity can 
be skipped in the AREM-GRFT processing.

\emph{Step 5: Perform constant false alarm rate (CFAR) detection under a given false alarm probability. }
\begin{equation}
\left| {G\left( {r_{s,i}}, {\dot r_{s,j}}, {v_{s,q}} \right)} \right| \mathop{\gtrless} \limits_{H_0}^{H_1} \eta 
\end{equation}
where $\eta$ is the CFAR detection threshold \cite{RFRFT-2014}. 
Use the amplitude of AREM-GRFT output as test statistic and perform CFAR detection to confirm targets. 
The adaptive detection threshold can be obtained by the reference cells 
in the output matrix under the given false alarm probability. 
If the test statistic is less than the detection threshold, 
there is no target to be declared. 
If the test statistic is larger than the detection threshold, 
a target is detected.

\emph{Step 6: Estimate target motion parameters. }

After the target is confirmed, 
the corresponding peak located in range-Doppler-speed domain can achieve the estimation of target motion parameters. 
Based on the initial range, initial radial velocity and speed corresponding to the coordinate of peak in range-Doppler-speed domain, 
the estimated parameters $\left( {{{\hat r}_0},{{\hat {\dot r}}_0},\hat v} \right)$ can be obtained. 
In addition, the trajectory of the moving target can be achieved by 
\begin{equation}
\hat r\left( {{t_m}} \right) = \sqrt{\hat r_0^2 + 2{{\hat r}_0}{{\hat {\dot r}}_0}{t_m} + {{\hat v}^2}t_m^2}
\end{equation}

\subsection{Computational complexity}\label{Computational Complexity}

\begin{table} [!t] \renewcommand{\arraystretch}{1.25} 
	\centering
	\caption{Computational complexity.}
	\label{Computational complexity}
	\setlength{\tabcolsep}{3mm}{
		\begin{tabular}{lcc}
			\hline
			\hline
			Methods         &   Computational complexity                                  &  Search dimension       \\
			\hline  		
			RFT             &   $O\left( {{N_r}{N_{\dot r}}M} \right)$                    &  2-D search             \\		
			
			KT-MFP          &   $O\left( {{N_k}{N_a}M{N_r}\log {}_2M} \right)$            &  2-D search             \\
			
			GRFT            &   $O\left( {{N_r}{N_{\dot r}}{N_a}{N_{\dot a}}M} \right)$   &  4-D search             \\
			
			Proposed        &   $O\left( {{N_r}{N_{\dot r}}{N_{v}}M} \right)$             &  3-D search             \\
			\hline
	\end{tabular}}
\end{table}

In this subsection, the computational complexity of the proposed method is discussed 
and compared against those of RFT \cite{RFT1-2011}, KT-MFP \cite{KT-MFP-2017}, and GRFT \cite{GRFT-2012}. 
The number of integrated pulses, 
searching range, 
searching radial velocity, 
searching speed, 
searching radial acceleration, 
searching radial jerk and searching fold factor 
are denoted by $M$, ${N_r}$, ${N_{\dot r}}$, ${N_{v}}$, ${N_a}$, ${N_{\dot a}}$ and ${N_k}$, respectively.

Table \ref{Computational complexity} gives the computational complexity of 
RFT \cite{RFT1-2011}, 
KT-MFP \cite{KT-MFP-2017}, 
GRFT \cite{GRFT-2012} 
and the proposed method. 
RFT requires two-dimensional (2-D) parameter search to correct the first-order RM caused by radial velocity, 
whose computational cost is $O\left( {{N_r}{N_{\dot r}}M} \right)$ \cite{RFT1-2011}. 
KT-MFP further eliminates the RM and DFM caused by radial acceleration with KT operation and matched filtering process. 
The computational complexity of KT operation is in the order of $ M^{2} N_{r} $ and 
the computational complexity of matched filtering process is in the order of $ {{N_k}{N_a}M{N_r}\log {}_2M} $. 
Hence, the computational burden of KT-MFP is $O\left( {{N_k}{N_a}M{N_r}\log {}_2M} \right)$ \cite{KT-MFP-2017}. 
As to the proposed method, its computational complexity is $O\left( {{N_r}{N_{\dot r}}{N_{v}}M} \right)$ 
because of three-dimensional (3-D) parameter search in range-Doppler-speed domain. 
GRFT involves four-dimensional (4-D) parameter search 
(i.e., search of range, radial velocity, radial acceleration and radial jerk), 
whose computational complexity is $O\left( {{N_r}{N_{\dot r}}{N_a}{N_{\dot a}}M} \right)$ \cite{GRFT-2012}.

In summary, the computational complexity of the proposed method is higher than that of RFT and KT-MFP, 
but lower than GRFT's. 
Normally, for the polynomial model based LTCI methods, 
extending the order of Taylor polynomial model could approximate the optimal integration gain, 
but it also means a sharp increase in computational complexity, 
which makes the polynomial model based LTCI methods face to 
the contradiction between integration gain and computational complexity. 
On the contrary, 
the proposed method can achieve better integration performance than 
RFT, KT-MFP and GRFT for LTCI problem of CCV target with feasible computational complexity, 
as will be illustrated in the numerical simulation experiments.

\begin{figure}[!t]  
	\centering 
	\includegraphics[scale= 0.85 ]{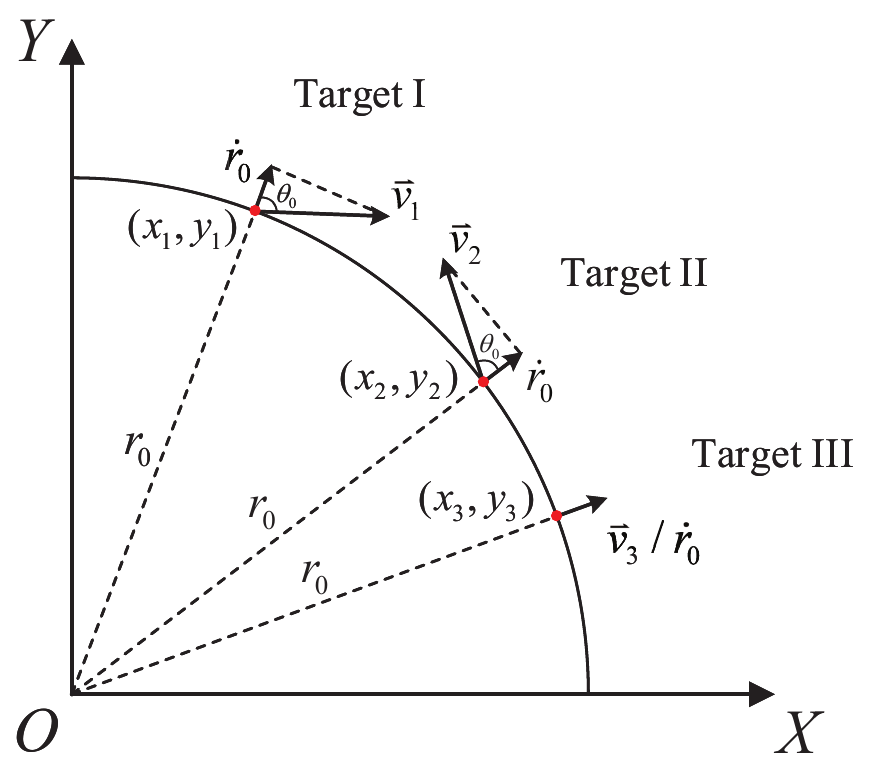}  
	\caption{Relationship of target motion parameters in Cartesian and range coordinates.}  
	\label{relationship}
\end{figure}

%%%%%%%%%%%%%%%%%%%%%%%%%%%%%%%%%%%%%%%%%%%%%%%%%%%%%%%%%%%%%%%%%%%%%%%%%%%%%%%%Numerical Examples
\section{Numerical simulation experiments}\label{Numerical Simulation Experiments}
In order to evaluate the performance of the proposed method, 
comprehensive experiments are presented in this section. 
Before the specification of the scenario parameters, 
the relationship of target motion parameters in Cartesian and range coordinates is discussed, 
as illustrated in Fig. \ref{relationship}. 
For a CCV target (e.g., Target I), 
its motion can be summarized by 
the initial Cartesian position $\left(x_{0}, y_{0}\right)$ and the Cartesian velocity vector $\vec{v}$. 
Given these parameters in Cartesian coordinates, 
the characteristic motion parameters in range coordinate of the target, 
initial range $r_{0}$, initial radial velocity $\dot{r}_{0}$, and speed $|\vec{v}|$, can be uniquely determined. 
On the other hand, 
given the set of initial target range $r_{0}$, initial radial velocity $\dot{r}_{0}$, and speed $|\vec{v}|$, 
there will be innumerable possible CCV targets, 
such as Target I and Target II in Fig. \ref{relationship}. 
Since Target I and Target II have the same initial range, initial radial velocity and speed, 
they cannot be distinguished by the proposed method and 
there will be only one peak in the range-Doppler-speed domain. 
When the absolute value of the target's radial velocity is equal to target's speed, 
the CCV motion is reduced to a constant radial velocity motion (see Target III in Fig. \ref{relationship}). 
In this case, 
both the existing polynomial model based LTCI method and the proposed method can achieve effective energy integration. 
However, the target moves with constant radial velocity is just a special case of CCV motion. 
Most CCV targets violate the constant radial velocity assumption and any polynomial models with finite orders. 
This problem becomes serious in the case of long integration time and target with high speed. 
In these cases, the transverse component of target velocity, i.e., the difference between radial velocity and velocity, 
cannot be ignored.

To verify the effectiveness and superiority of the proposed method, 
simulation experiments are performed for both constant radial velocity motion and common CCV motion, 
where the cases with long integration time and high speed are explored. 
Meanwhile, the multi-target integration performance and detection performance of the proposed method are also evaluated. 
Several typical LTCI methods are used for comparison, 
including first-order polynomial model based method (i.e., RFT) \cite{RFT1-2011}, 
second-order polynomial model based method (i.e., KT-MFP) \cite{KT-MFP-2017}, 
and high-order polynomial model based method (i.e., GRFT) \cite{GRFT-2012}.

Note that the PC gain of the pulse-compressed signal is normalized in the simulation experiments, 
so that the value of the output amplitude of the target reflects the effective integrated pulse number. 
If a method can provide an integration result whose amplitude is consistent with the pulse number, 
it means that the method achieves the ideal integration gain.

The radar parameters are listed in Table \ref{Table_Radar_Parameters}.

\begin{table} [!t] \renewcommand{\arraystretch}{1.2} 
	\centering
	\caption{Radar parameters.}
	\label{Table_Radar_Parameters}
	\setlength{\tabcolsep}{5mm}{
		\begin{tabular}{ll}
			\hline
			\hline
			Parameters                  &    Value (Unit)       \\
			\hline
			Carrier frequency           &    1.5 (${\rm{GHz}}$)   \\	
			
			Pulse duration time         &    10 (${\rm{us}}$)     \\		
			
			Signal bandwidth            &    20 (${\rm{MHz}}$)    \\
			
			Range sampling frequency    &    50 (${\rm{MHz}}$)    \\					
			
			Pulse repetition frequency  &    200 (${\rm{Hz}}$)    \\
			
			Pulse number                &    500                \\
			\hline
	\end{tabular}}
\end{table}

\subsection{Coherent integration for a single target}\label{Coherent Integration for a Single Target}
In this subsection, three scenarios 
are considered to evaluate the performance of the proposed AREM-GRFT method: 
1) constant radial velocity motion, which is a special case of CCV motion; 
2) CCV motion for long integration time; 
3) CCV motion for target with high speed.

\subsubsection{Performance in case of constant radial velocity motion}\label{Performance in case of Constant Radial Velocity Motion}
When the absolute value of the target's radial velocity is equal to target's speed, 
the CCV motion is reduced to a constant radial velocity motion. 
In this scenario, the target parameters are set as: 
initial range $r_{0}$ $=$ 25 $\mathrm{km}$, 
initial radial velocity $\dot{r}_{0}$ $=$ 800 $\mathrm{m} / \mathrm{s}$, 
speed $v$ $=$ 800 $\mathrm{m} / \mathrm{s}$, 
and SNR after PC is 6 $\rm{dB}$. 
The radar parameters are the same as those in Table \ref{Table_Radar_Parameters}.

\begin{figure}[htbp]
	\centering
	\subfigure[]{
		\includegraphics[width= 4.5 cm]{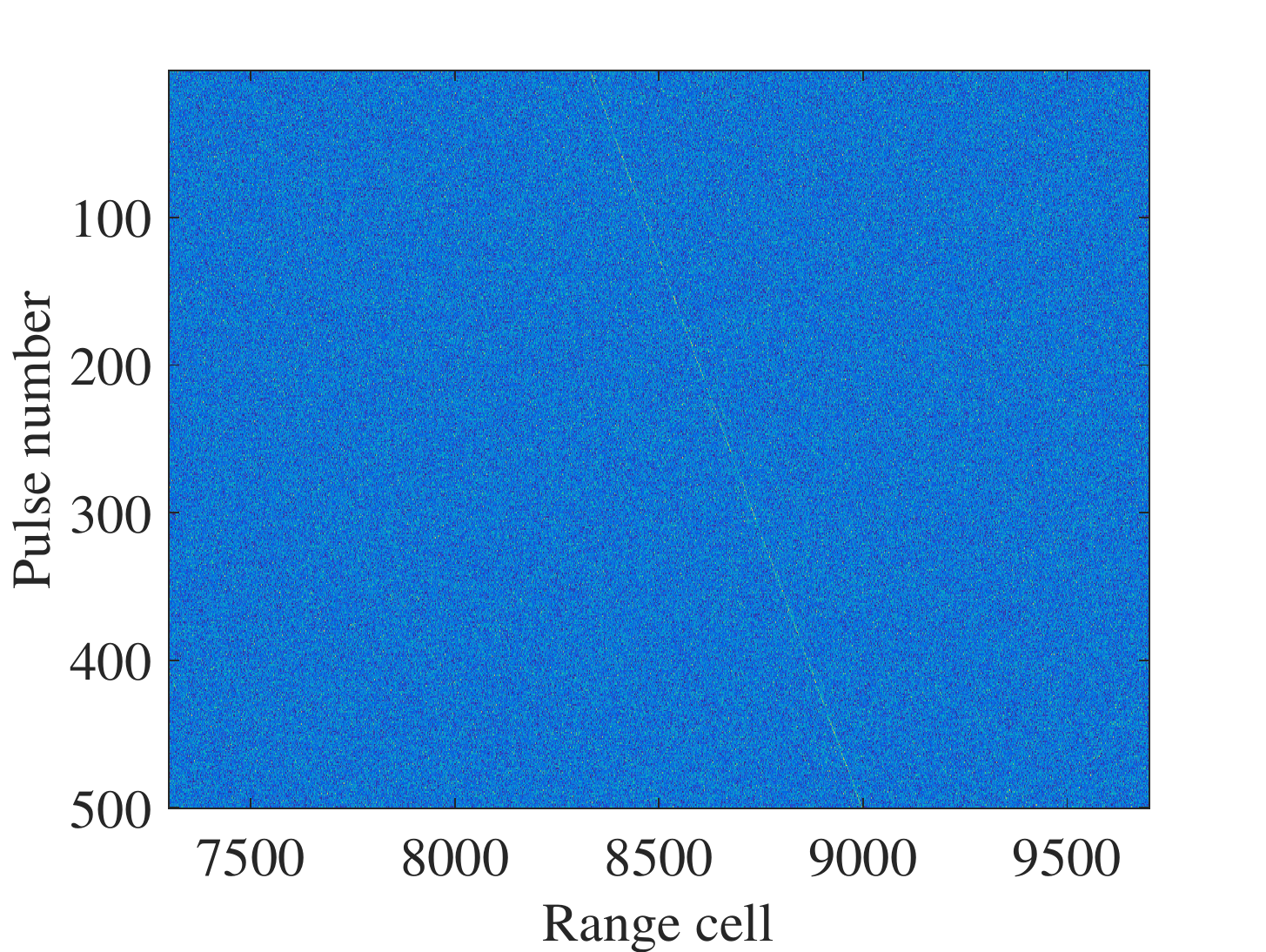} \label{PC_radial_motion} }
	\subfigure[]{
		\includegraphics[width= 4.5 cm]{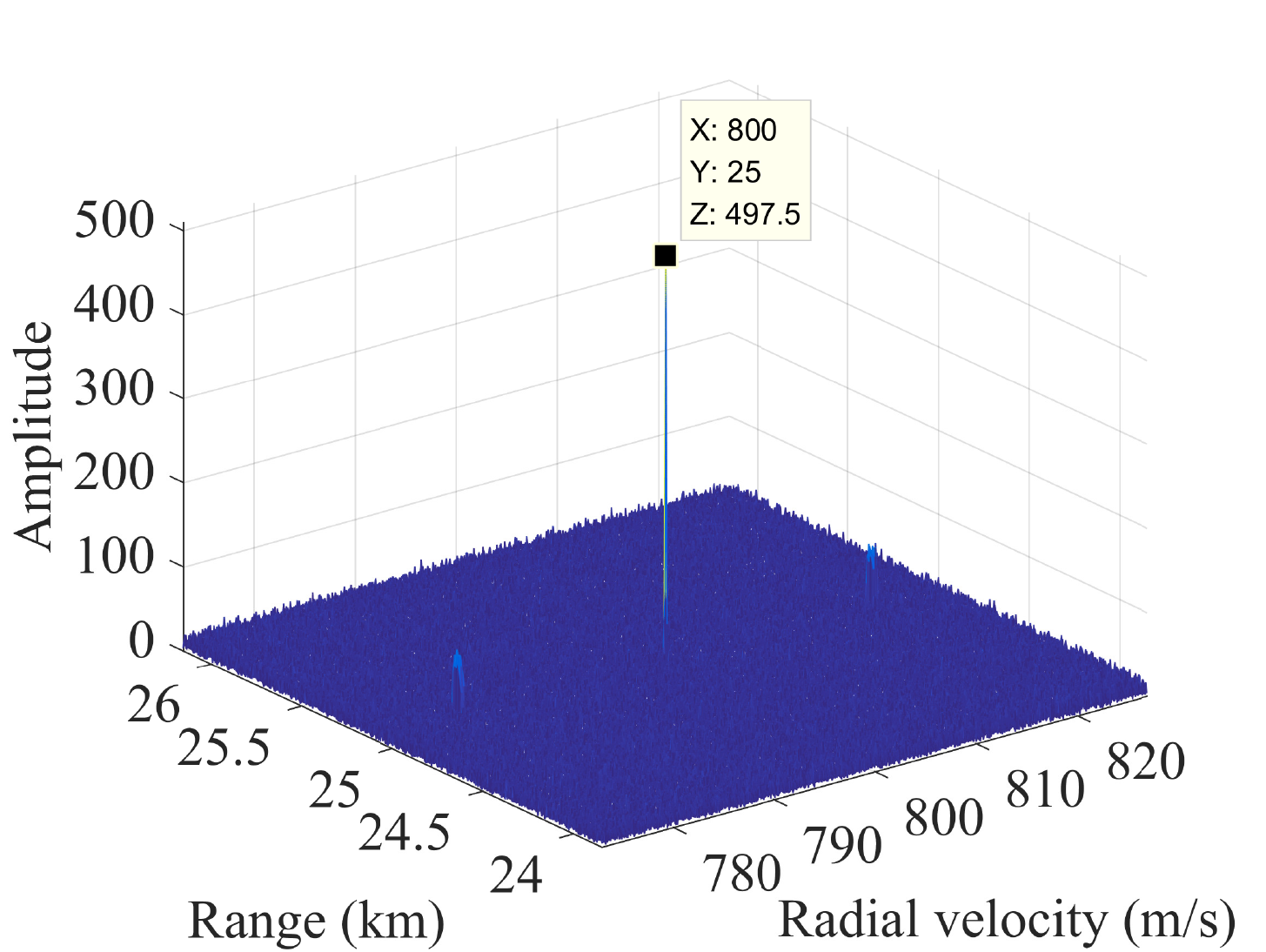} \label{RFT_radial_motion} }
	\subfigure[]{
		\includegraphics[width= 4.5 cm]{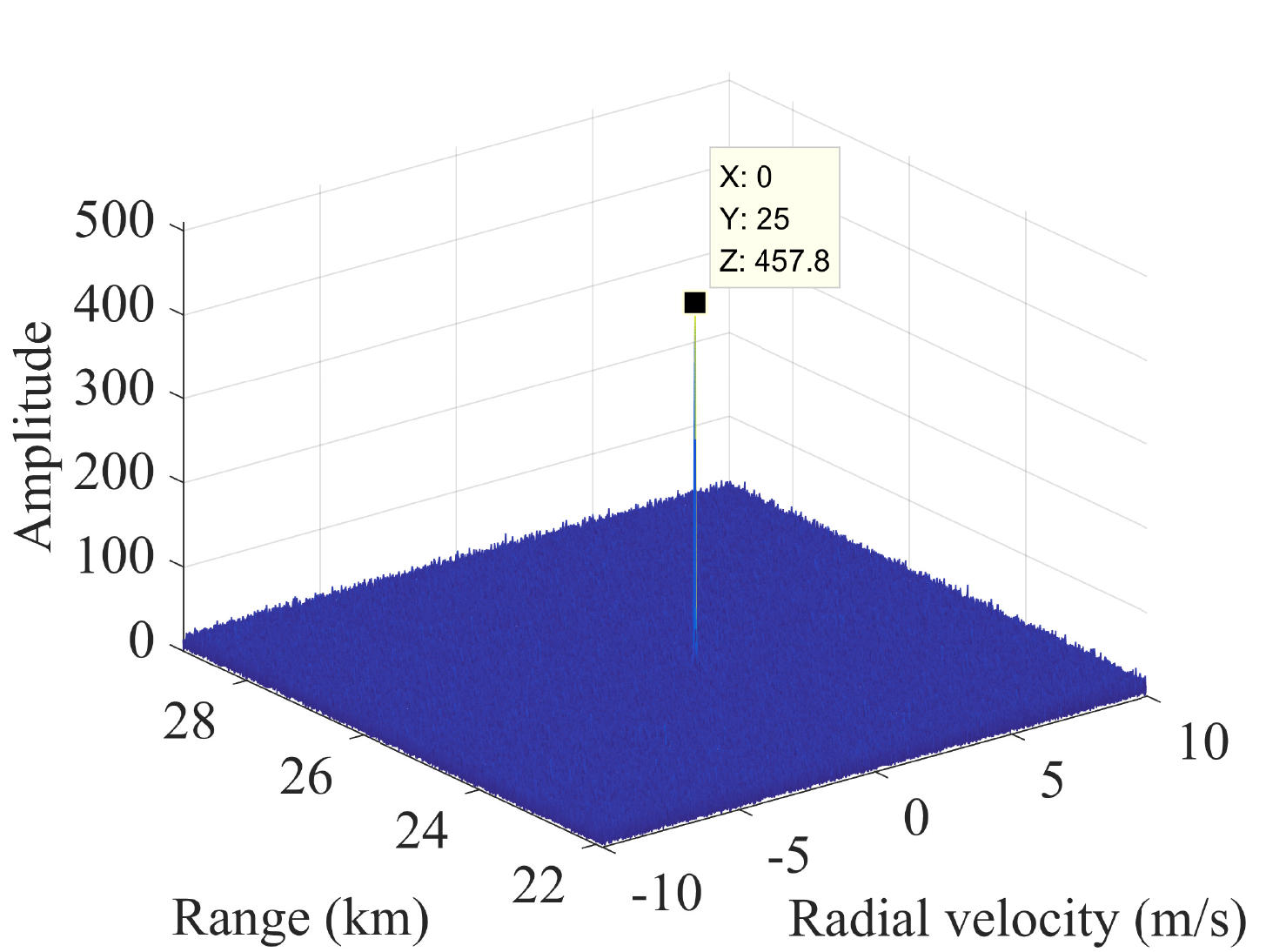} \label{KT_MFP_radial_motion} }
	\subfigure[]{
		\includegraphics[width= 4.5 cm]{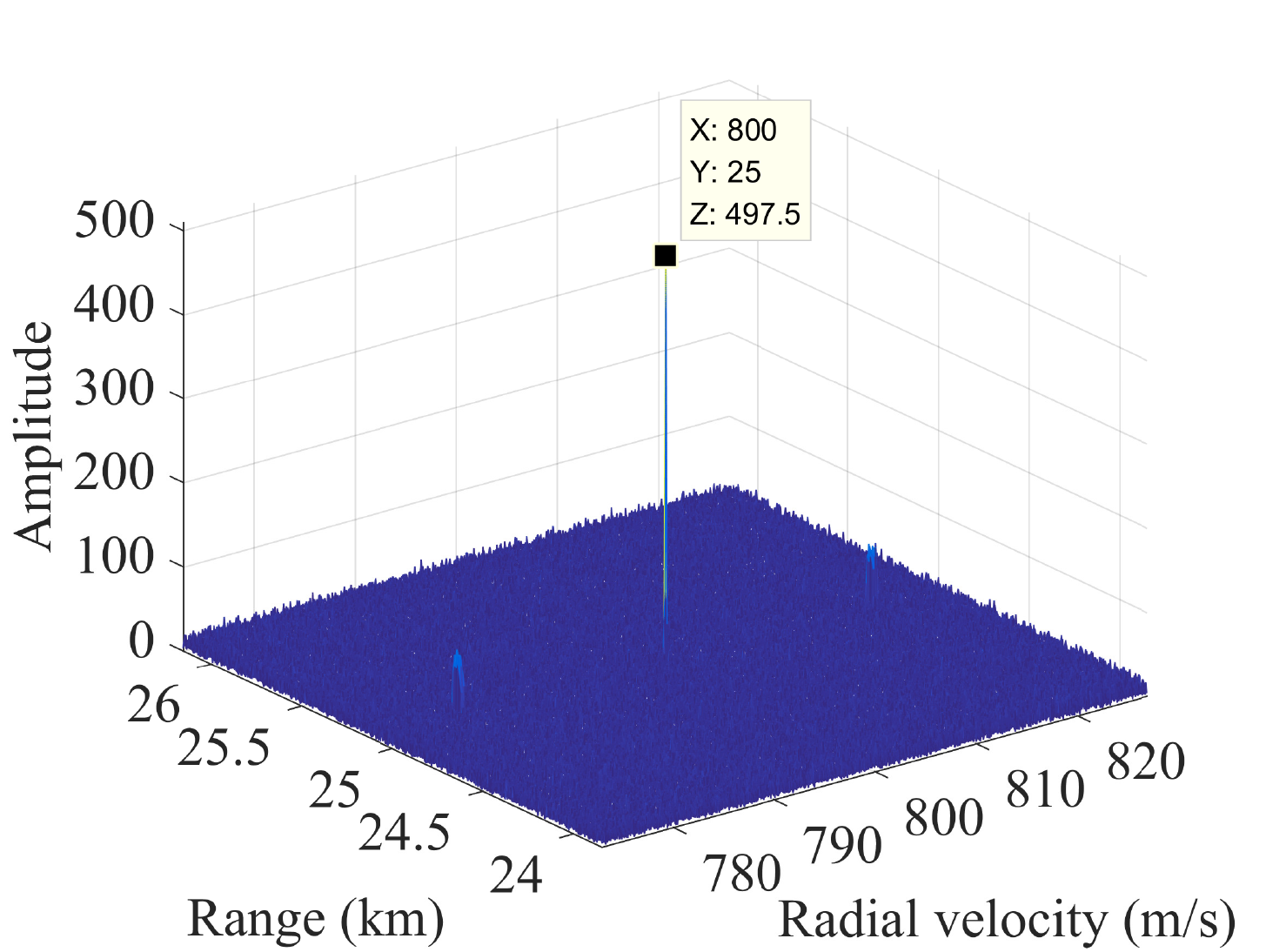} \label{GRFT_radial_motion} } 	
	\subfigure[]{
		\includegraphics[width= 4.5 cm]{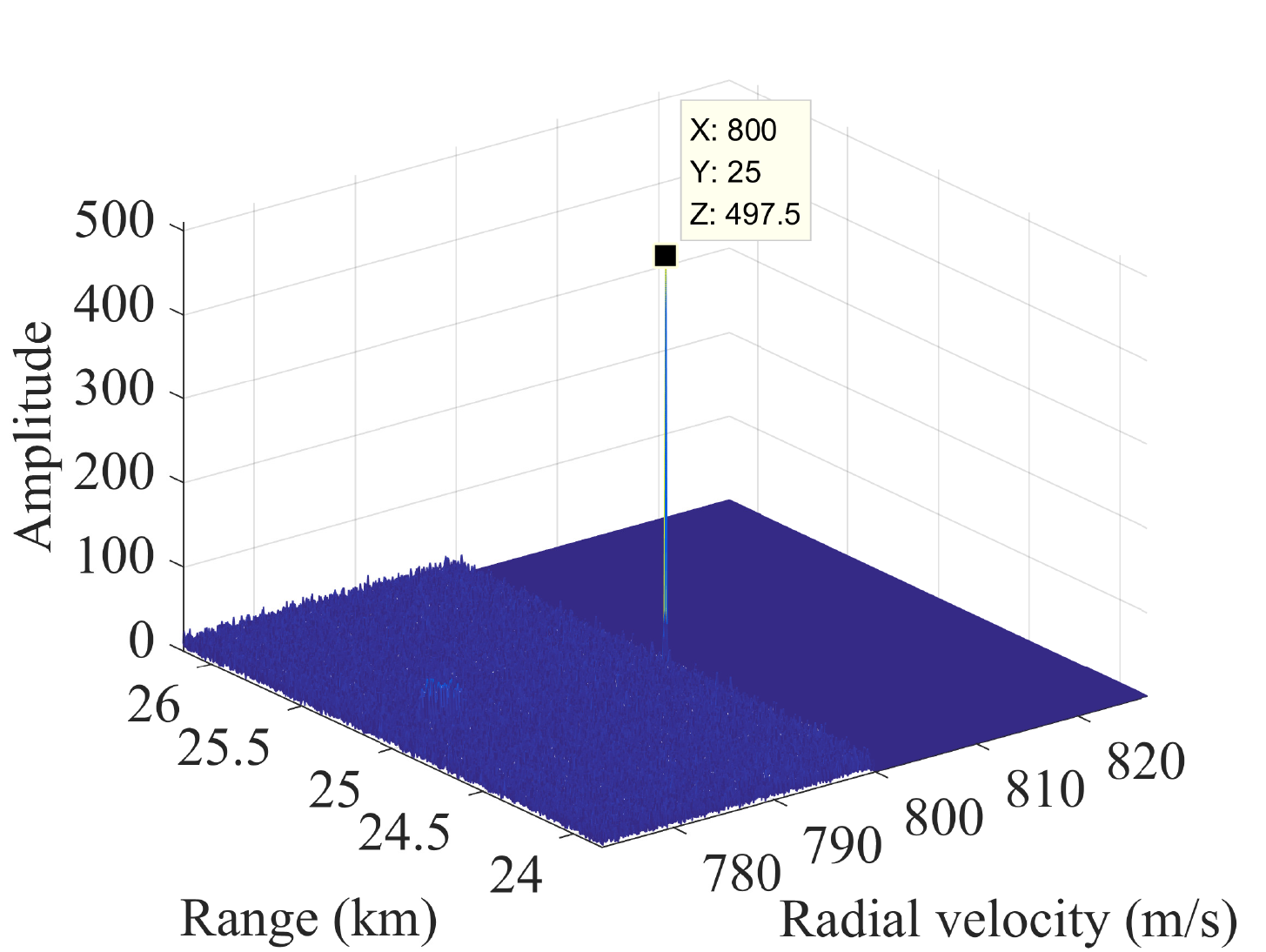} \label{VVslice_radial_motion} }
	\subfigure[]{
		\includegraphics[width= 4.5 cm]{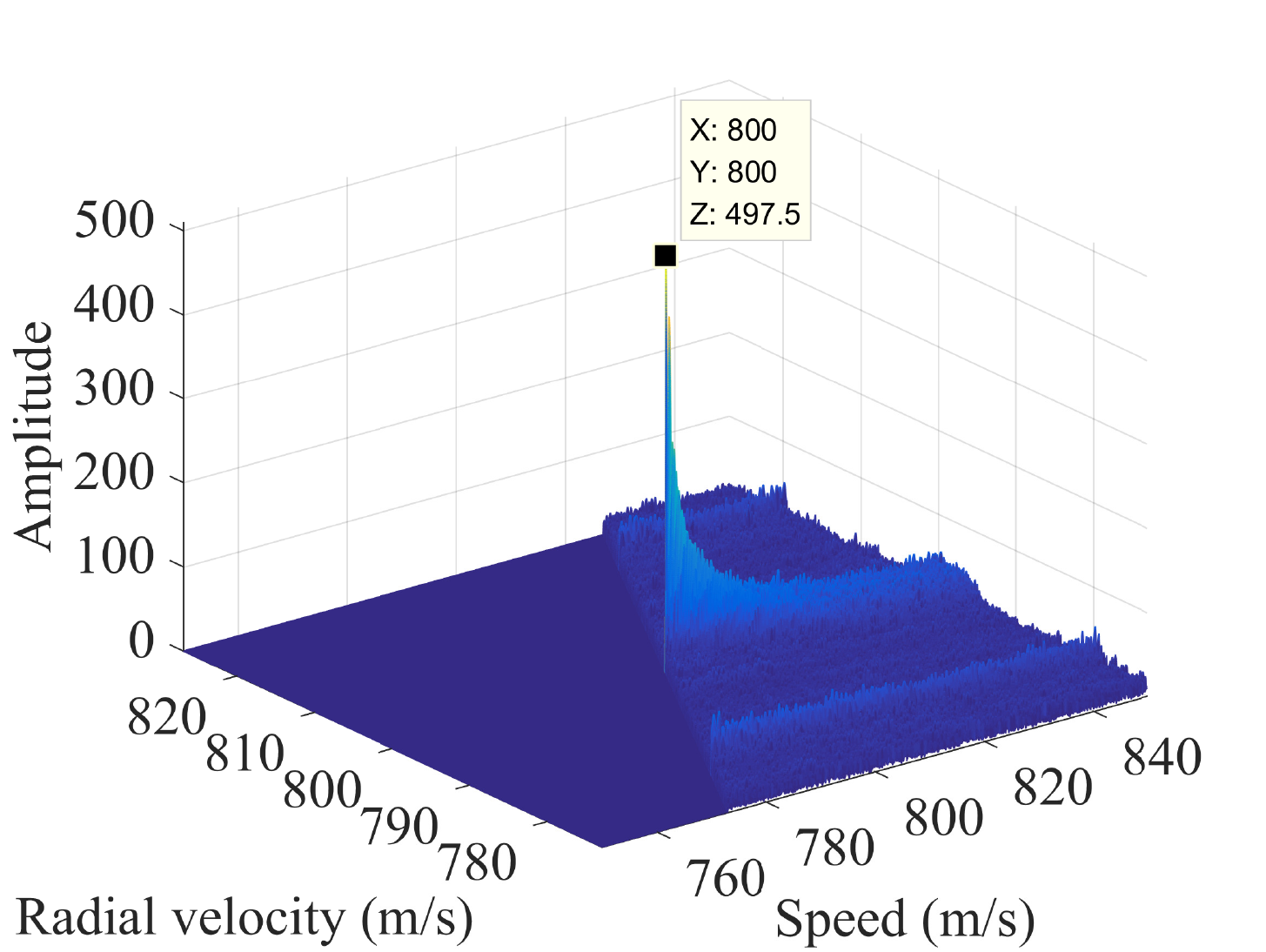} \label{Rslice_radial_motion} }
	\subfigure[]{
		\includegraphics[width= 4.5 cm]{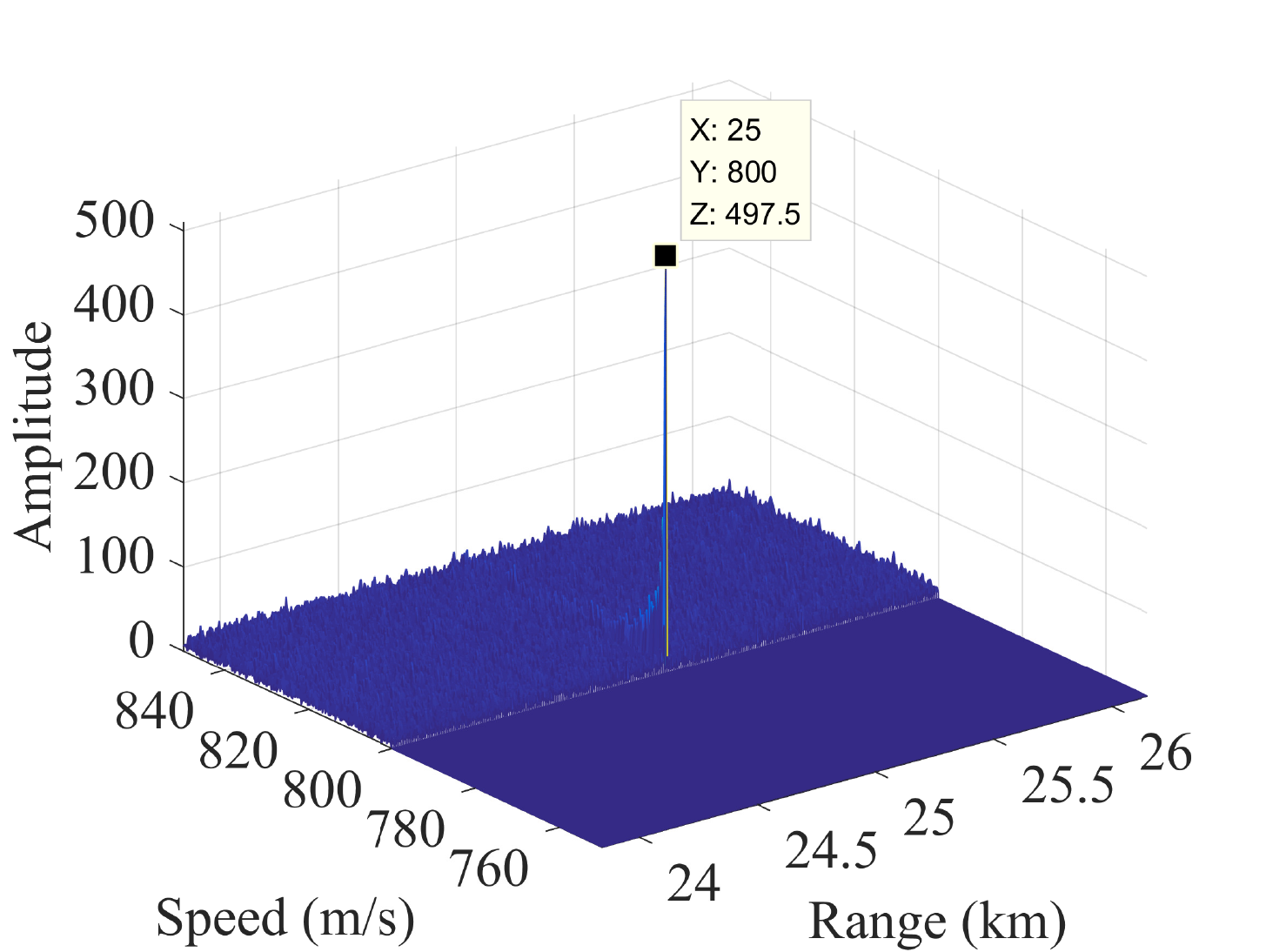} \label{Vslice_radial_motion} }
	\caption{Simulation results in case of constant radial velocity motion. 
		(a) Result after pulse compression. 
		(b) RFT. 
		(c) KT-MFP. 
		(d) GRFT. 
		(e) Range-Doppler slice of the AREM-GRFT at speed of 800 $\mathrm{m} / \mathrm{s}$. 
		(f) Doppler-speed slice of the AREM-GRFT at range of 25 ${\rm{km}}$. 
		(g) Speed-range slice of the AREM-GRFT at radial velocity of 800 $\mathrm{m} / \mathrm{s}$.} 	
	\label{Simulation results in case of constant radial velocity motion.}
\end{figure}

The corresponding simulation results are shown in Fig. \ref{Simulation results in case of constant radial velocity motion.}. 
Fig. \ref{PC_radial_motion} illustrates the pulse compression result. 
The integration results of RFT, KT-MFP, and GRFT are shown in Fig. \ref{RFT_radial_motion}--Fig. \ref{GRFT_radial_motion}, respectively. 
The integration results of the proposed method in the slice of range-Doppler, Doppler-speed, and speed-range at the target position 
are given in Fig. \ref{VVslice_radial_motion}--Fig. \ref{Vslice_radial_motion}, respectively. 
Note that the searching speed should be greater than 
the absolute value of the searching radial velocity in each case for the proposed method. 
The unreasonable cases are skipped in the AREM-GRFT processing. 
It can be seen from Fig. \ref{Simulation results in case of constant radial velocity motion.} that 
all the methods obtain well-focused results. 
The RFT, GRFT, and the proposed method produce almost ideal integration in the case of 
target with constant radial velocity motion. 
This is because the signal model used in all these methods match the constant radial velocity motion.

However, CCV targets usually move in a different direction than the radar line of sight. 
When the transverse component cannot be ignored 
(that is, the difference between the absolute value of radial velocity and speed is large) \cite{RFT1-2011}, 
the approximate polynomial motion models cannot match the CCV motion well. 
Moreover, the model mismatch will be enlarged in the case of 
long integration time and high speed, 
further resulting in performance degradation. 
In contrast, the proposed method is able to achieve effective LTCI 
without limitation of integration time and target speed, 
since signal extraction and compensation are performed according to an accurate range model. 
In the following, another two scenarios are considered to further verify the effectiveness of the proposed method.

\subsubsection{Performance in case of long integration time}\label{AREM-GRFT of Long Integration Time}
To evaluate the performance of the proposed method for long integration time, 
the pulse number in the radar parameters changes to 800, 
and other radar parameters are the same as those in Table \ref{Table_Radar_Parameters}. 
The parameters of the target with CCV motion are set as: 
initial range $r_{0}$ $=$ 25 $\mathrm{km}$, 
initial radial velocity $\dot{r}_{0}$ $=$ 60 $\mathrm{m} / \mathrm{s}$, 
speed $v$ $=$ 800 $\mathrm{m} / \mathrm{s}$, 
and SNR after PC is 6 $\rm{dB}$.

\begin{figure}[htbp]
	\centering
	\subfigure[]{
		\includegraphics[width= 4.5 cm]{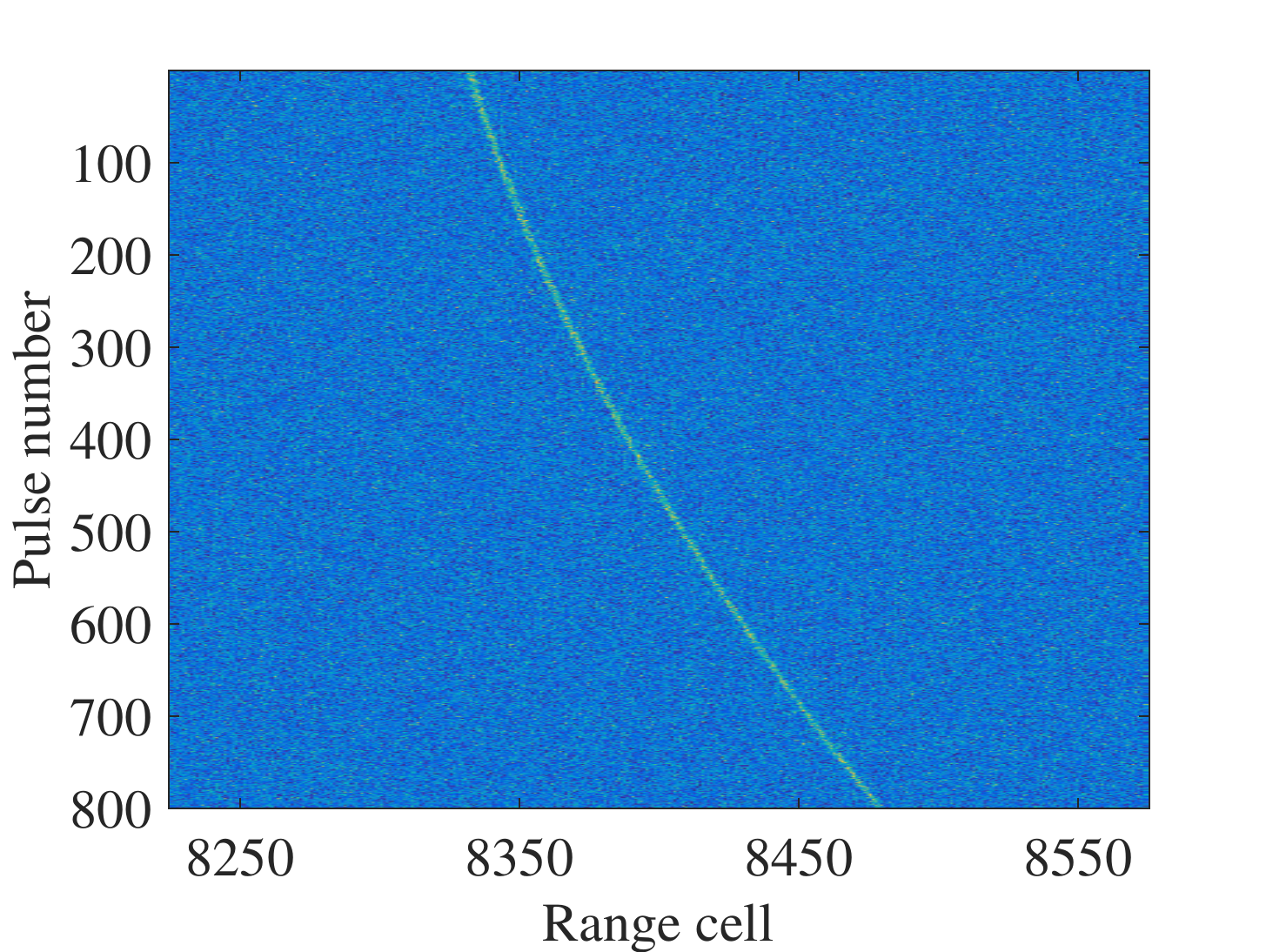} \label{PC2} }
	\subfigure[]{
		\includegraphics[width= 4.5 cm]{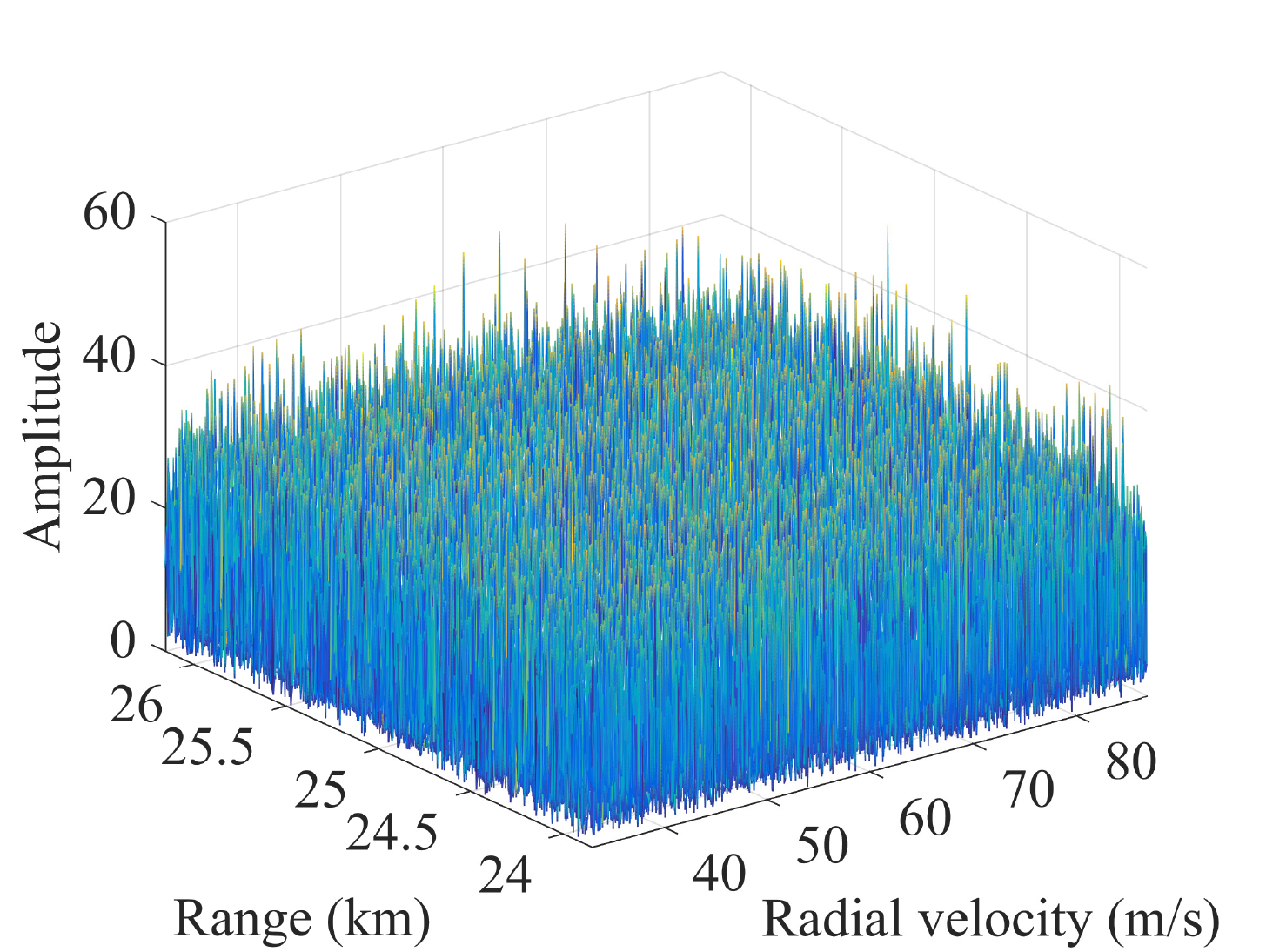} \label{RFT2} }
	\subfigure[]{
		\includegraphics[width= 4.5 cm]{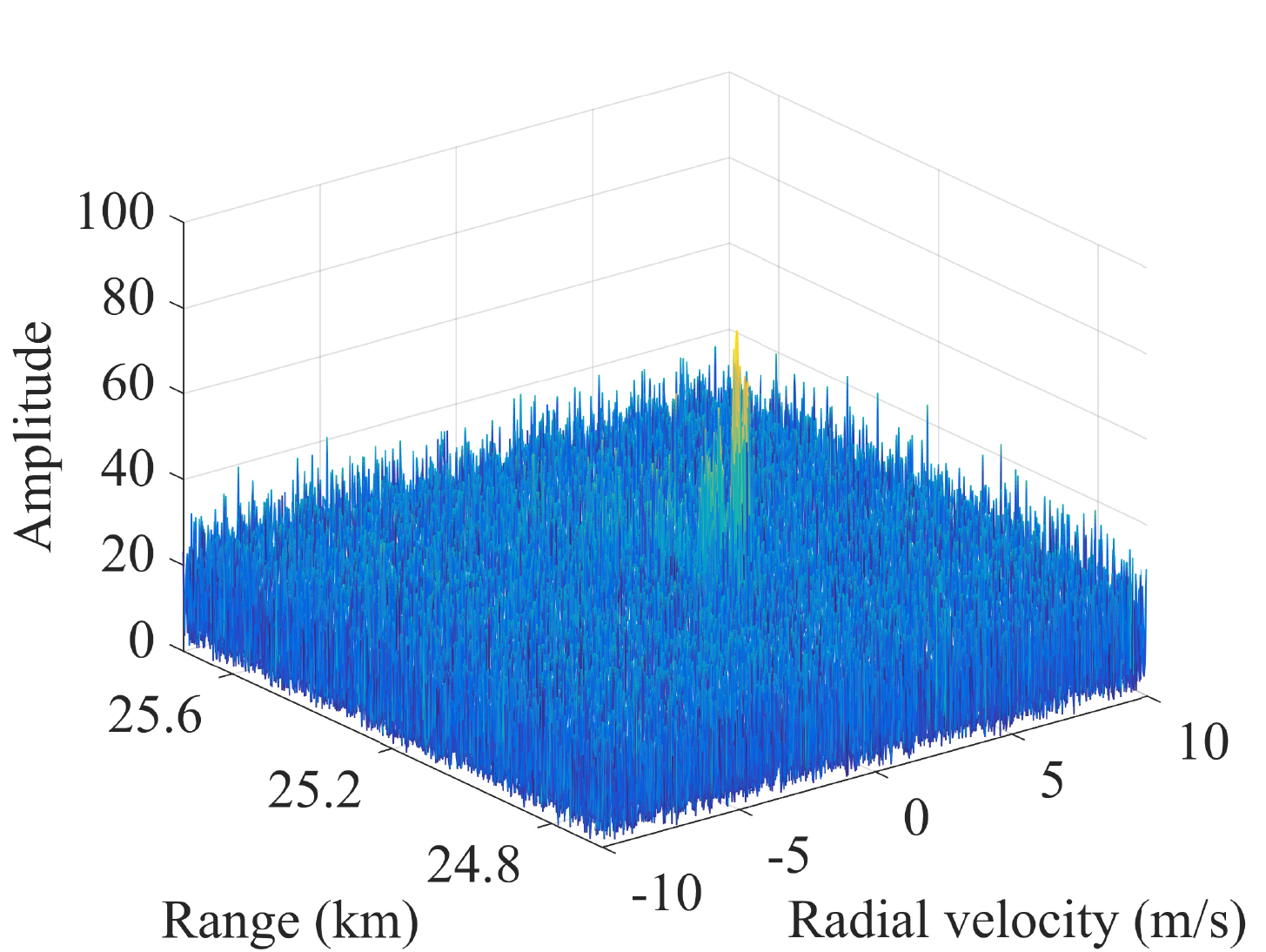} \label{KT_MFP2} }
	\subfigure[]{
		\includegraphics[width= 4.5 cm]{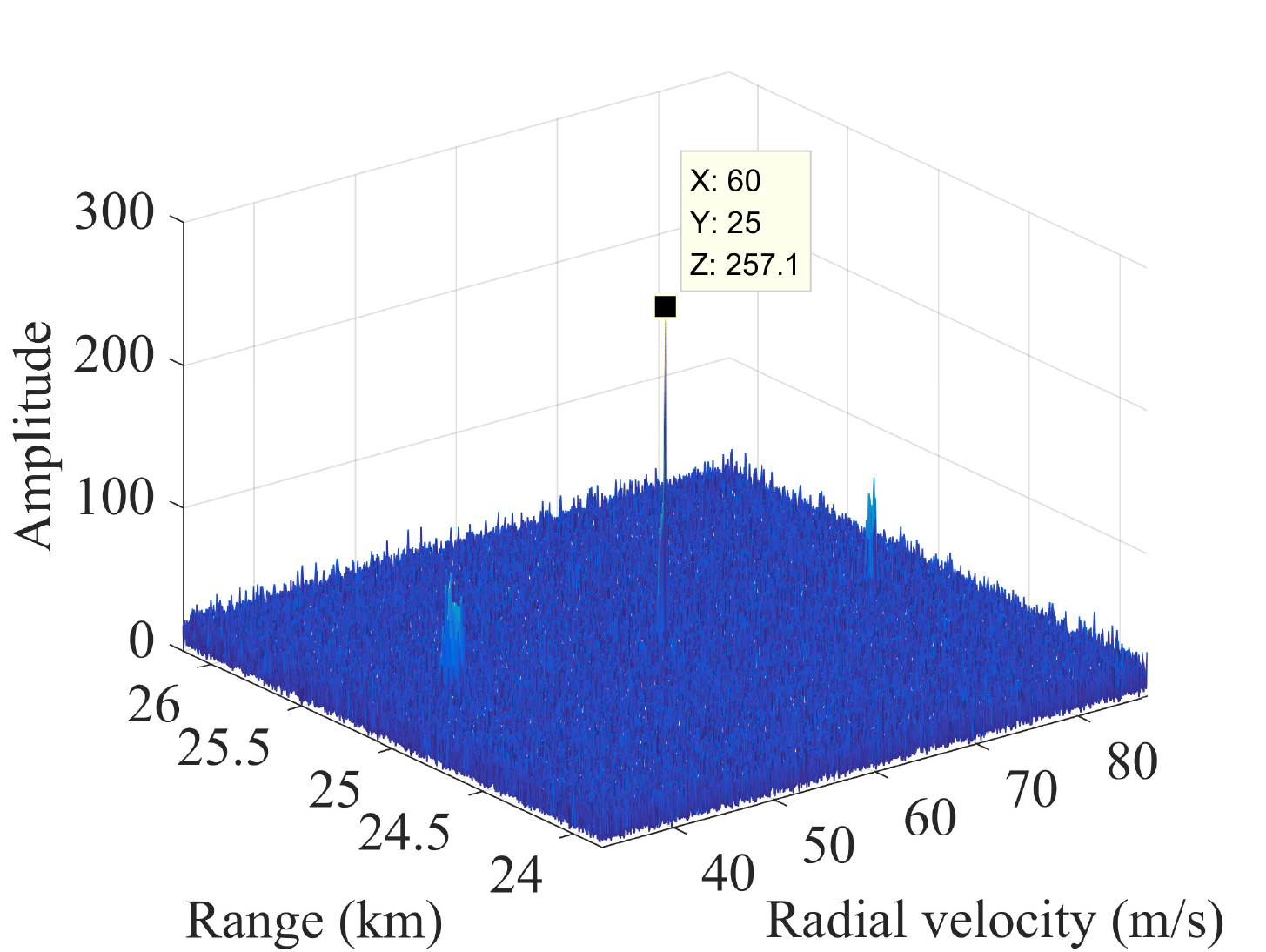} \label{GRFT2} }	
	\subfigure[]{
		\includegraphics[width= 4.5 cm]{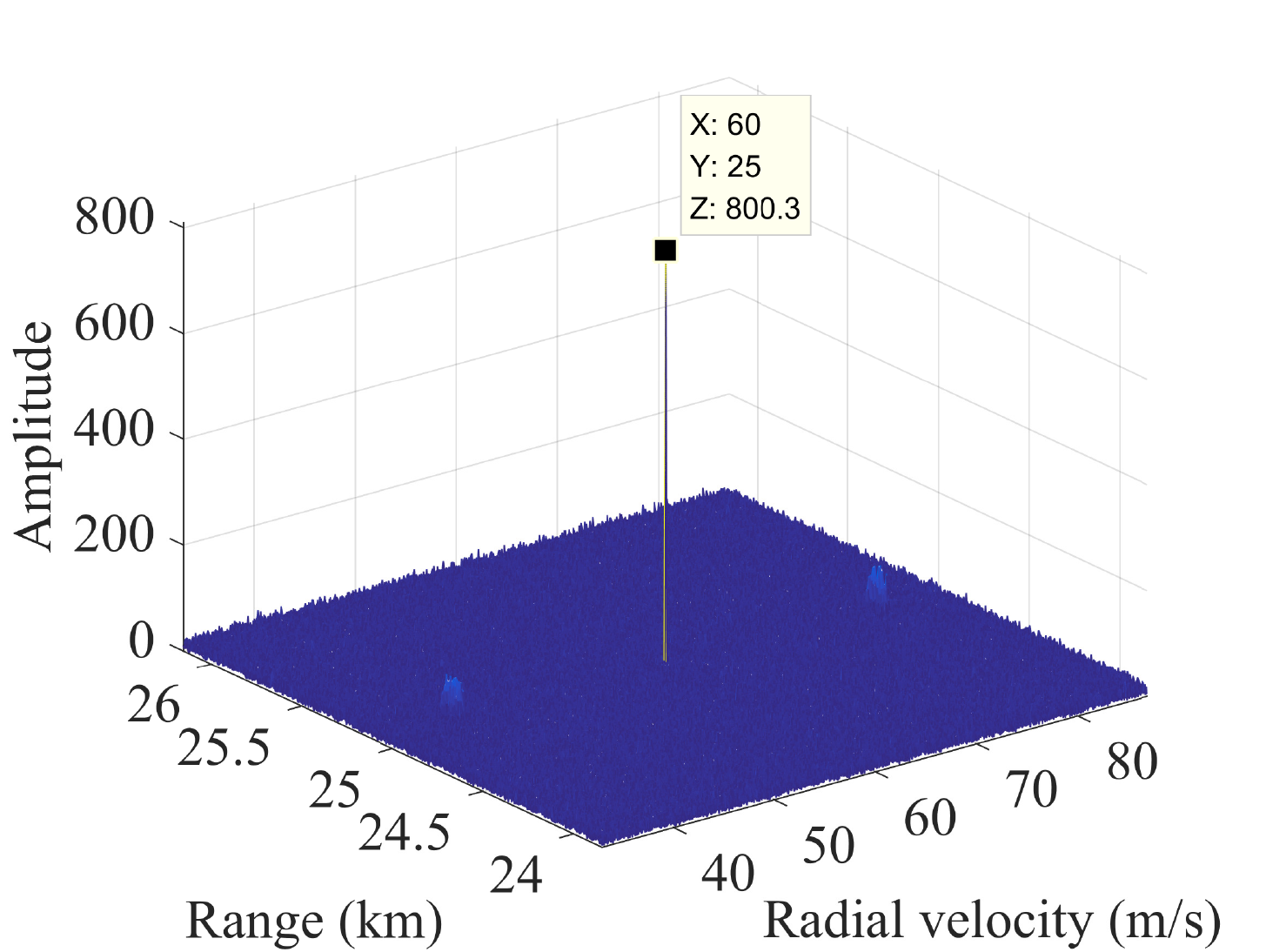} \label{VV_slice2} }	
	\subfigure[]{
		\includegraphics[width= 4.5 cm]{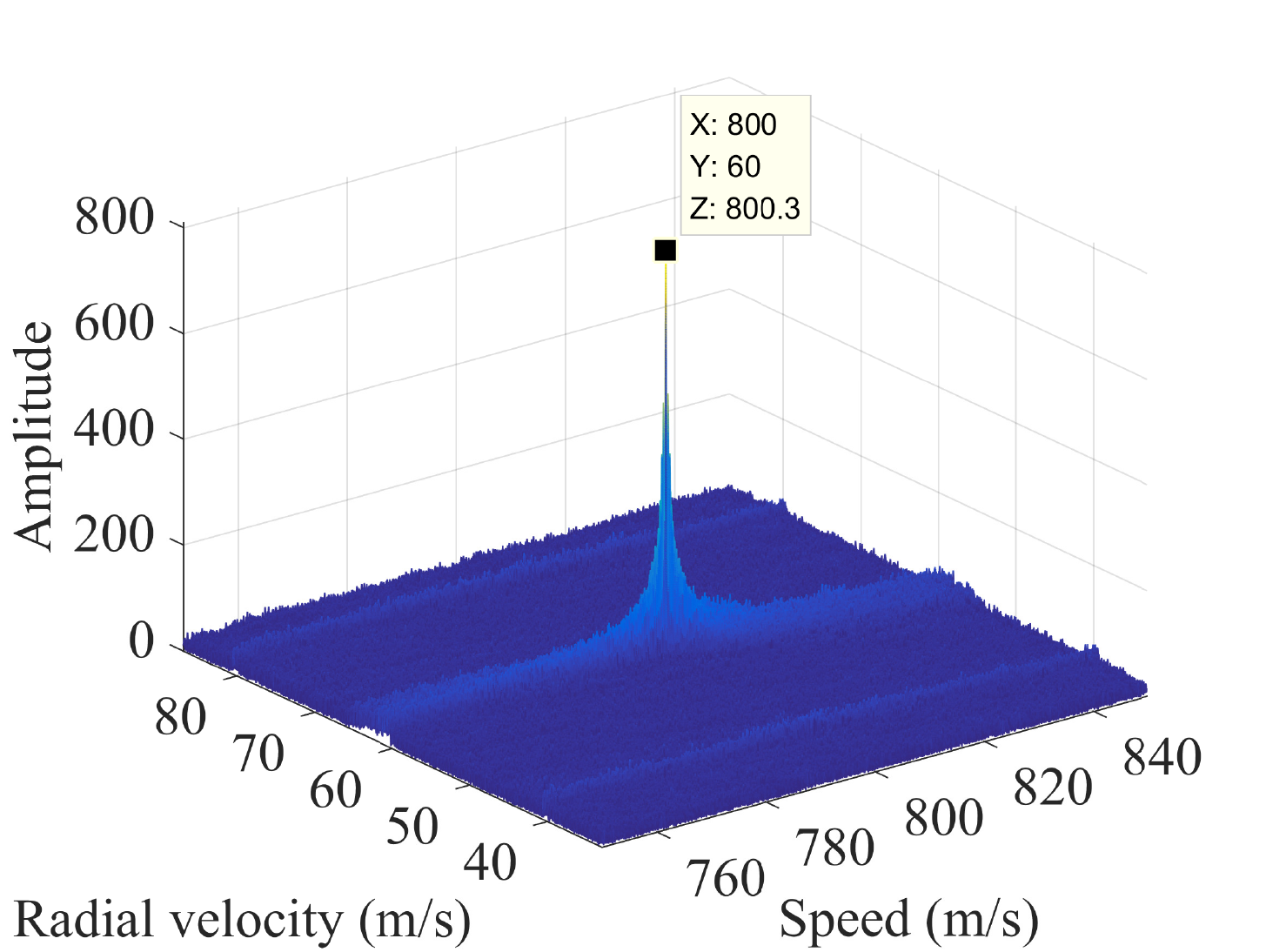} \label{R_slice2} }
	\subfigure[]{
		\includegraphics[width= 4.5 cm]{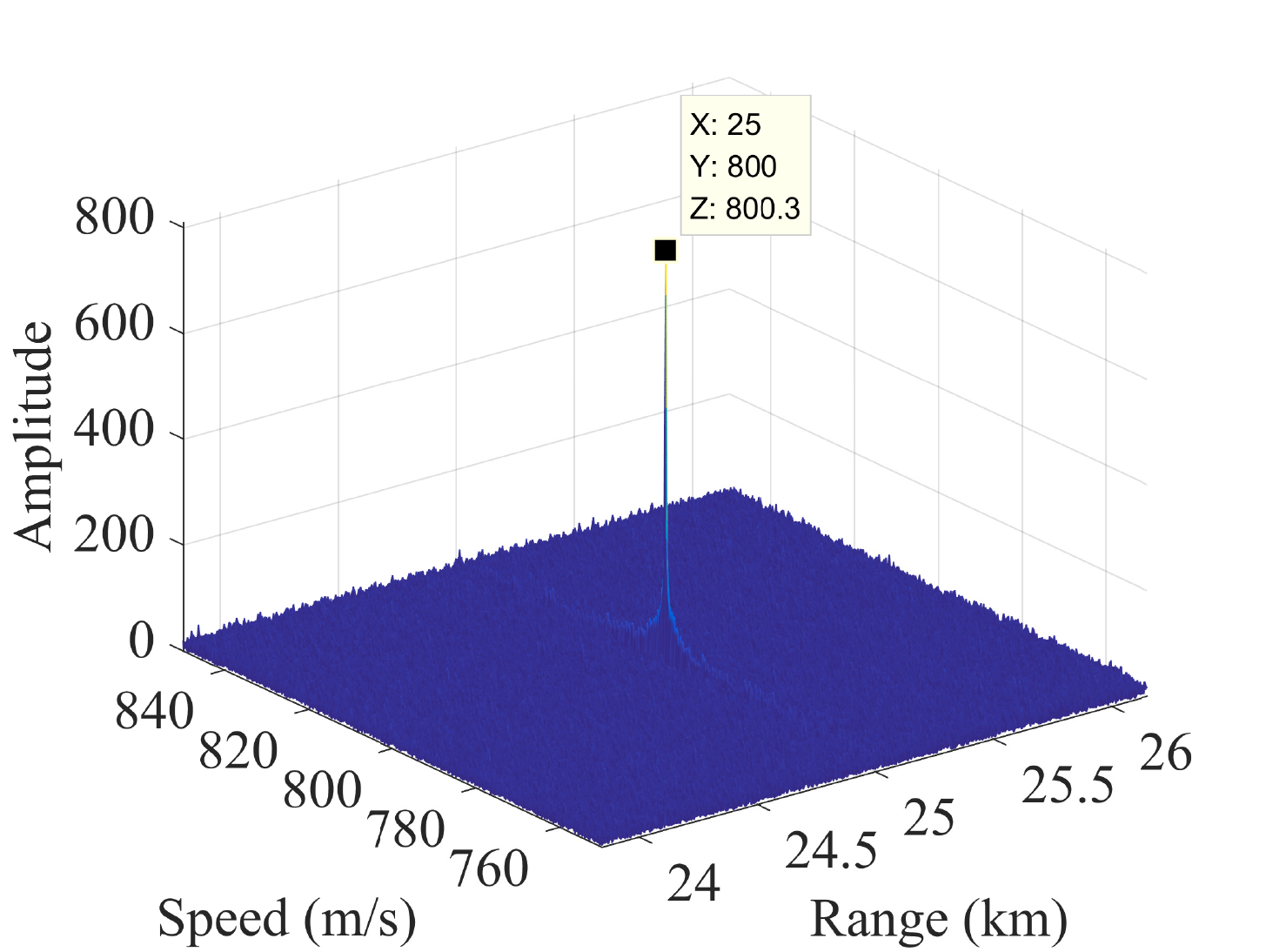} \label{V_slice2} }	
	\caption{Simulation results in case of long integration time. 
		(a) Result after pulse compression. 
		(b) RFT. 
		(c) KT-MFP. 
		(d) GRFT. 
		(e) Range-Doppler slice of the AREM-GRFT at speed of 800 $\mathrm{m} / \mathrm{s}$. 
		(f) Doppler-speed slice of the AREM-GRFT at range of 25 ${\rm{km}}$. 
		(g) Speed-range slice of the AREM-GRFT at radial velocity of 60 $\mathrm{m} / \mathrm{s}$.}
	\label{Simulation results in case of long integration time.}
\end{figure}

The simulation results in case of long integration time are illustrated 
in Fig. \ref{Simulation results in case of long integration time.}. 
It can be seen from the result of signal after pulse compression in Fig. \ref{PC2} that 
the trajectory of the target is highly nonlinear, 
which implies that the RM caused by the comprehensive effect of range, radial velocity, and speed is complex. 
The integration results of RFT, KT-MFP, and GRFT are shown in Fig. \ref{RFT2}--Fig. \ref{GRFT2}, respectively. 
The integration results of the proposed AREM-GRFT method 
in the slice of range-Doppler, Doppler-speed, and speed-range are given 
in Fig. \ref{VV_slice2}--Fig. \ref{V_slice2}, respectively.

It can be seen from Fig. \ref{RFT2} that 
the target energy resulted by the RFT is unfocused. 
This is because the RFT method could only correct the linear RM and 
has no capability of dealing with nonlinear RM and complex DFM. 
The result of KT-MFP is slightly better than that of RFT, 
and the GRFT is superior to the KT-MFP, as illustrated in Fig. \ref{KT_MFP2} and Fig. \ref{GRFT2}. 
It is because that KT-MFP method considers second-order RM and first-order DFM, and GRFT method further 
considers third-order RM and second-order DFM during the integration period. 
Normally, the higher the order of the polynomial model considered, 
the better the integration performance. 
However, the target energy resulted by the KT-MFP method still spreads over different range and Doppler cells, 
and the amplitude (i.e., effective integrated pulse number) of the GRFT method in Fig. \ref{GRFT2} is 257.1, 
which indicates that the coherent integration gain of GRFT is about 5 ${\rm{dB}}$ less than the ideal integration gain (i.e., 800). 
The integration results of KT-MFP method and GRFT method are not well-focused, 
since the CCV motion is highly nonlinear in long integration time. 
In summary, for the long integration time, 
these LTCI methods (i.e., RFT, KT-MFP and GRFT) cannot effectively eliminate 
the highly nonlinear RM and complex DFM caused by CCV motion.

On the contrary, it can be seen from Fig. \ref{VV_slice2}--Fig. \ref{V_slice2} that 
the proposed AREM-GRFT method can provide an effective integrated amplitude of 800.3 
and achieve the ideal integration gain, 
thanks to the range evolving model accurately matching the CCV motion. 
In addition, the estimated parameters (i.e., corresponding to the position of the peak) are consistent with the actual motion parameters. 
Note that the speed parameter can also be observed from the integration result of the AREM-GRFT method in addition to the range and Doppler parameters. 
This extra measurement is beneficial to tracking filter, data association and further applications.

\begin{figure}[htbp]
	\centering
	\subfigure[]{
		\includegraphics[width= 4.5 cm]{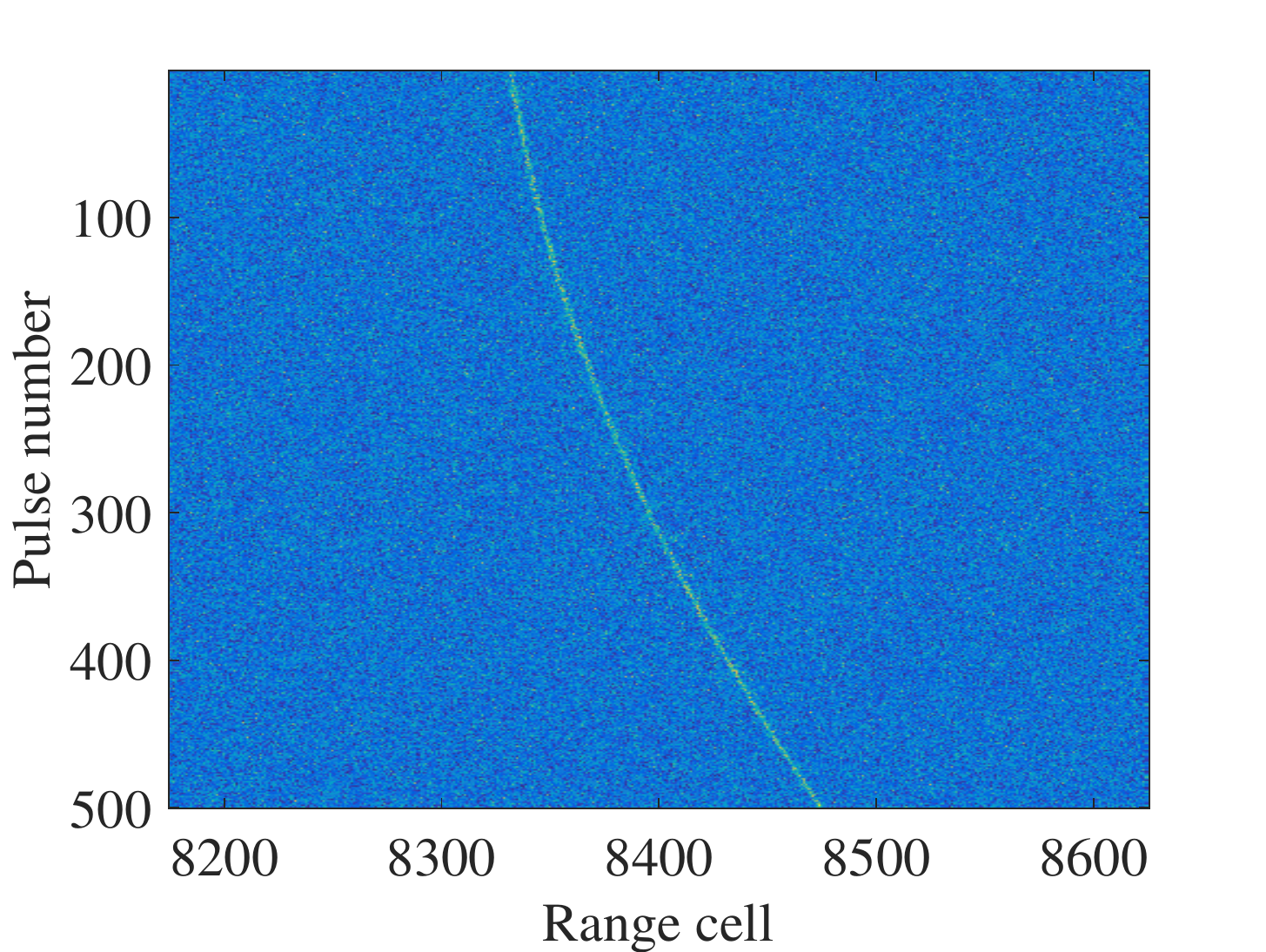} \label{PC4} }
	\subfigure[]{
		\includegraphics[width= 4.5 cm]{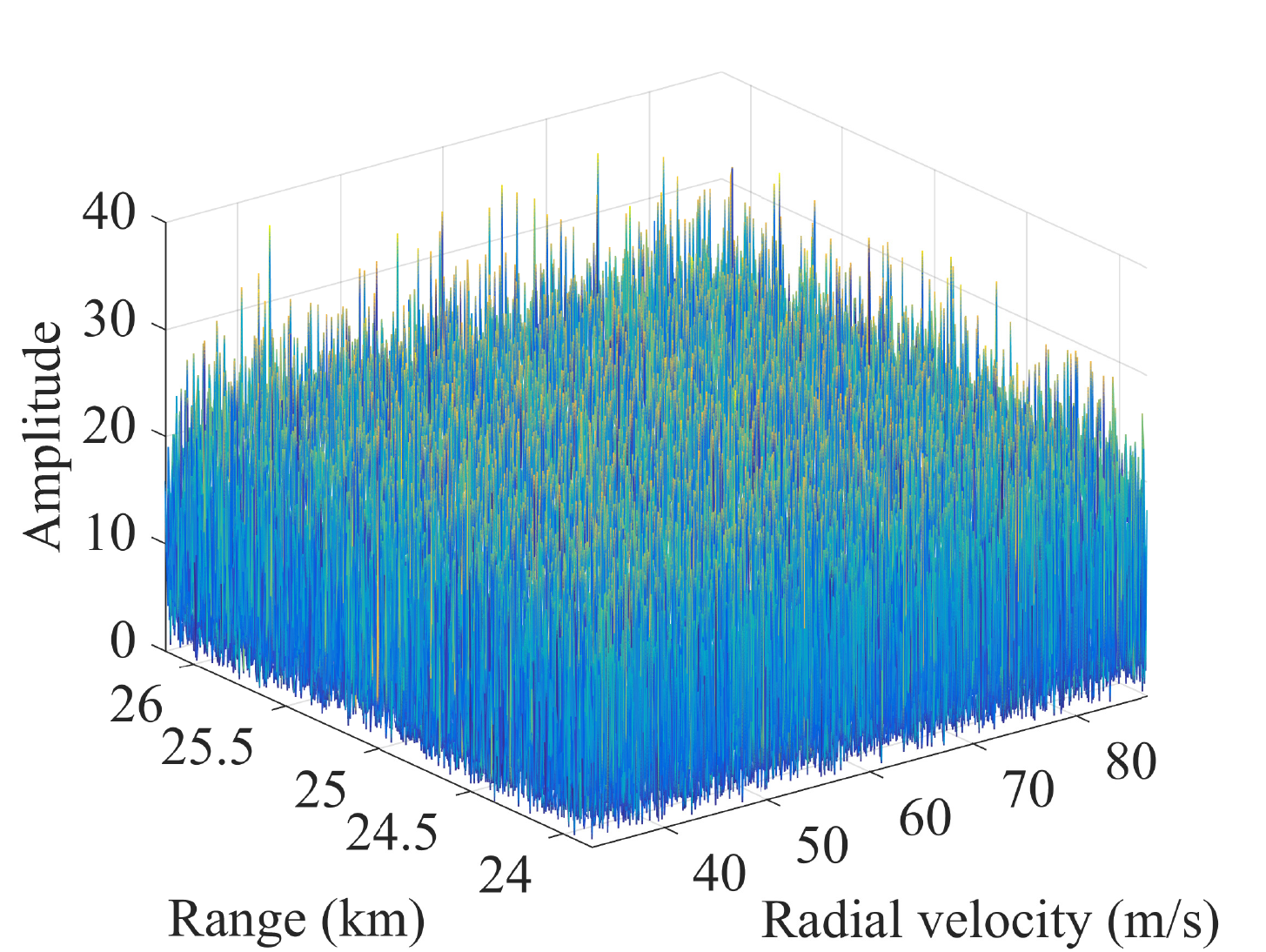} \label{RFT4} }
	\subfigure[]{
		\includegraphics[width= 4.5 cm]{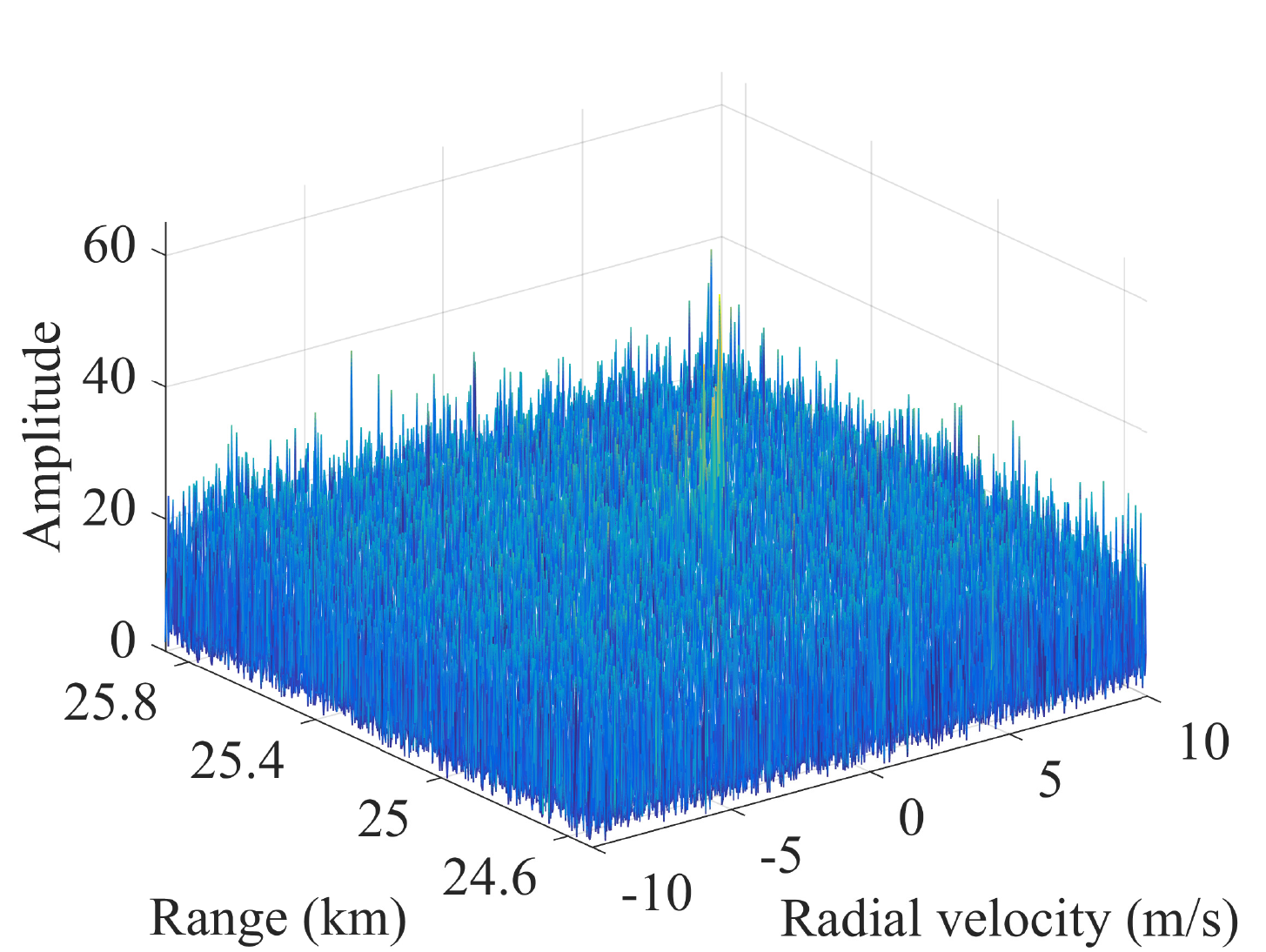} \label{KT_MFP4} }
	\subfigure[]{
		\includegraphics[width= 4.5 cm]{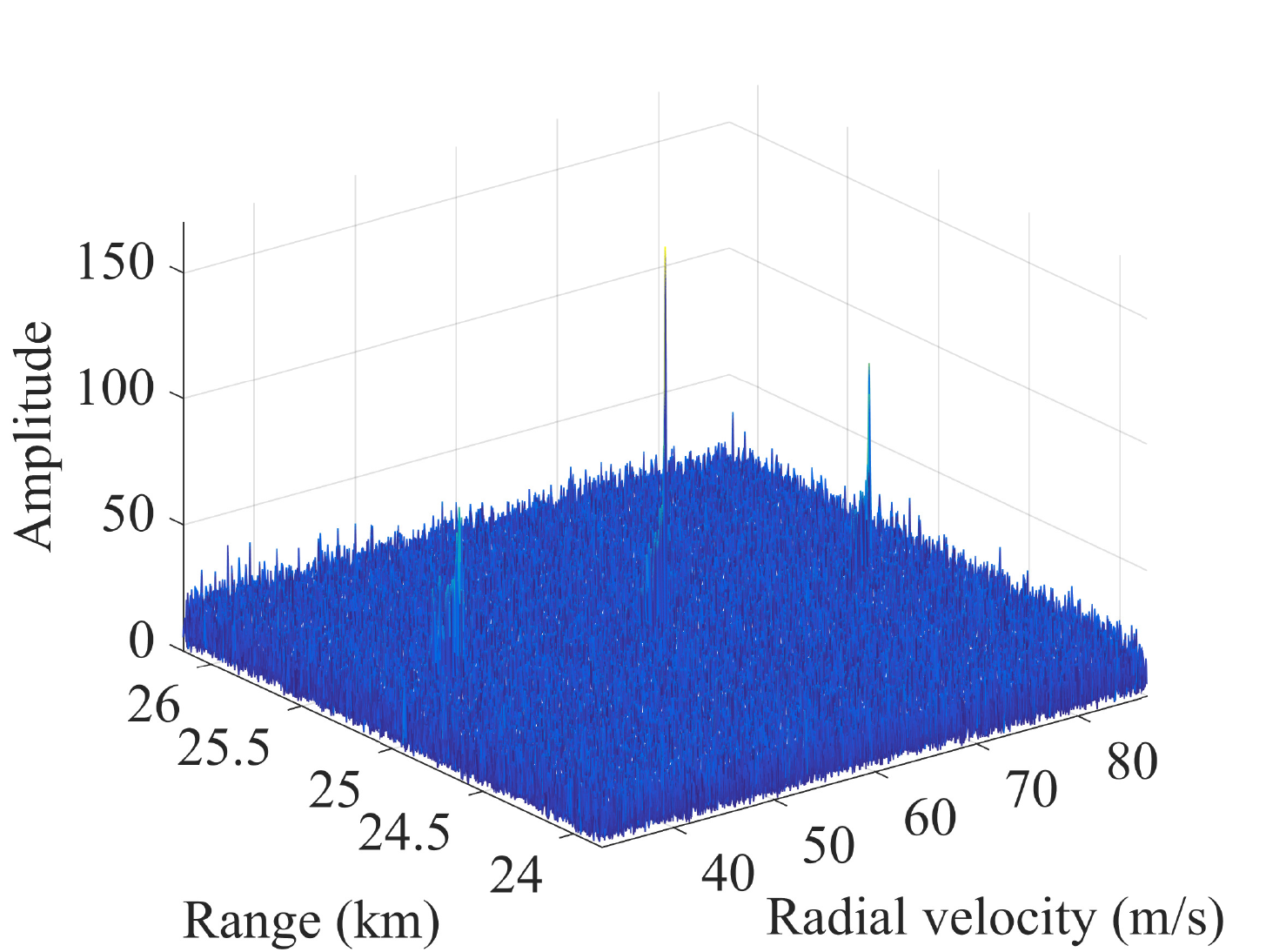} \label{GRFT4} }
	\subfigure[]{
		\includegraphics[width= 4.5 cm]{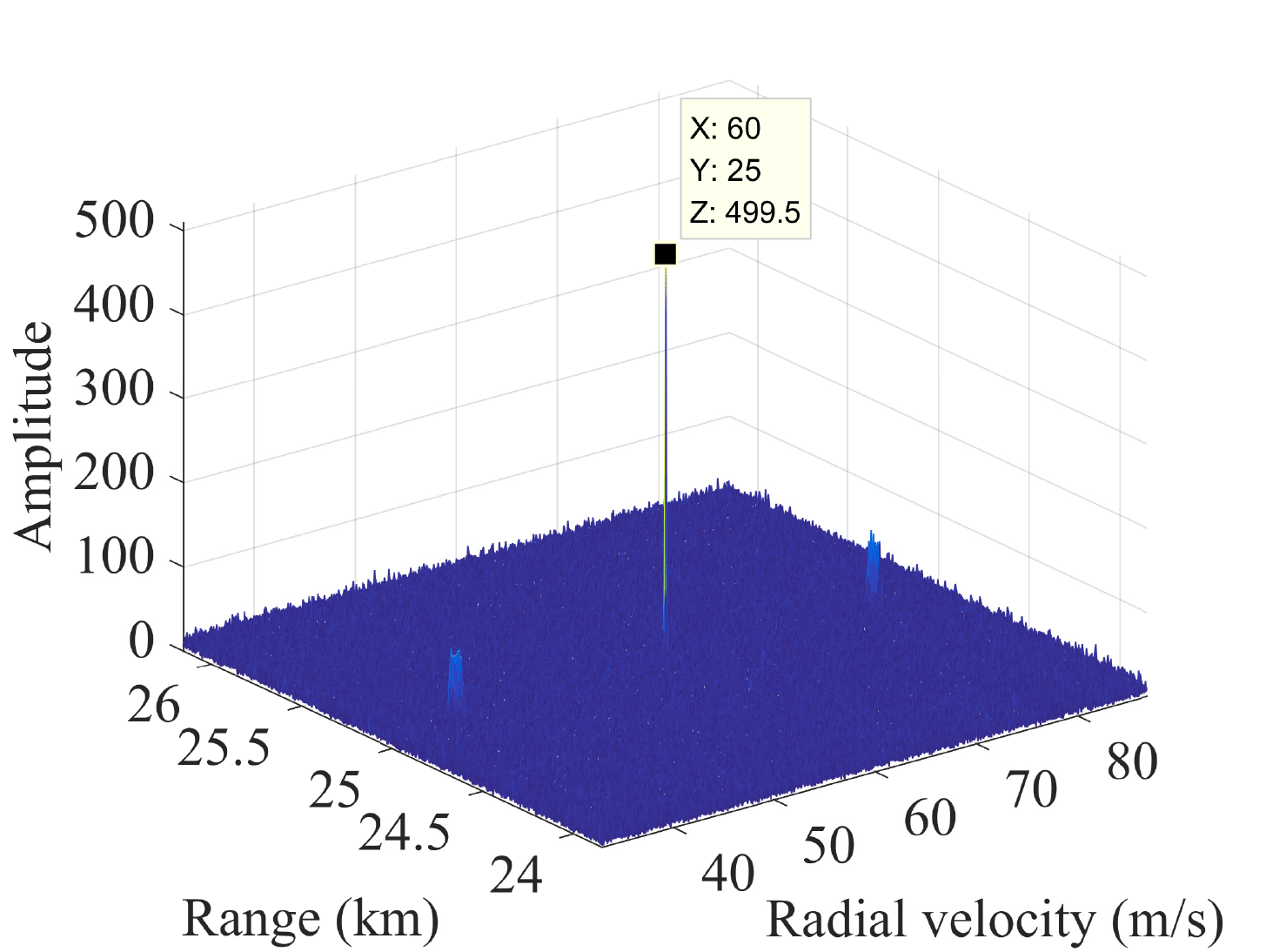} \label{VV_slice4} }
	\subfigure[]{
		\includegraphics[width= 4.5 cm]{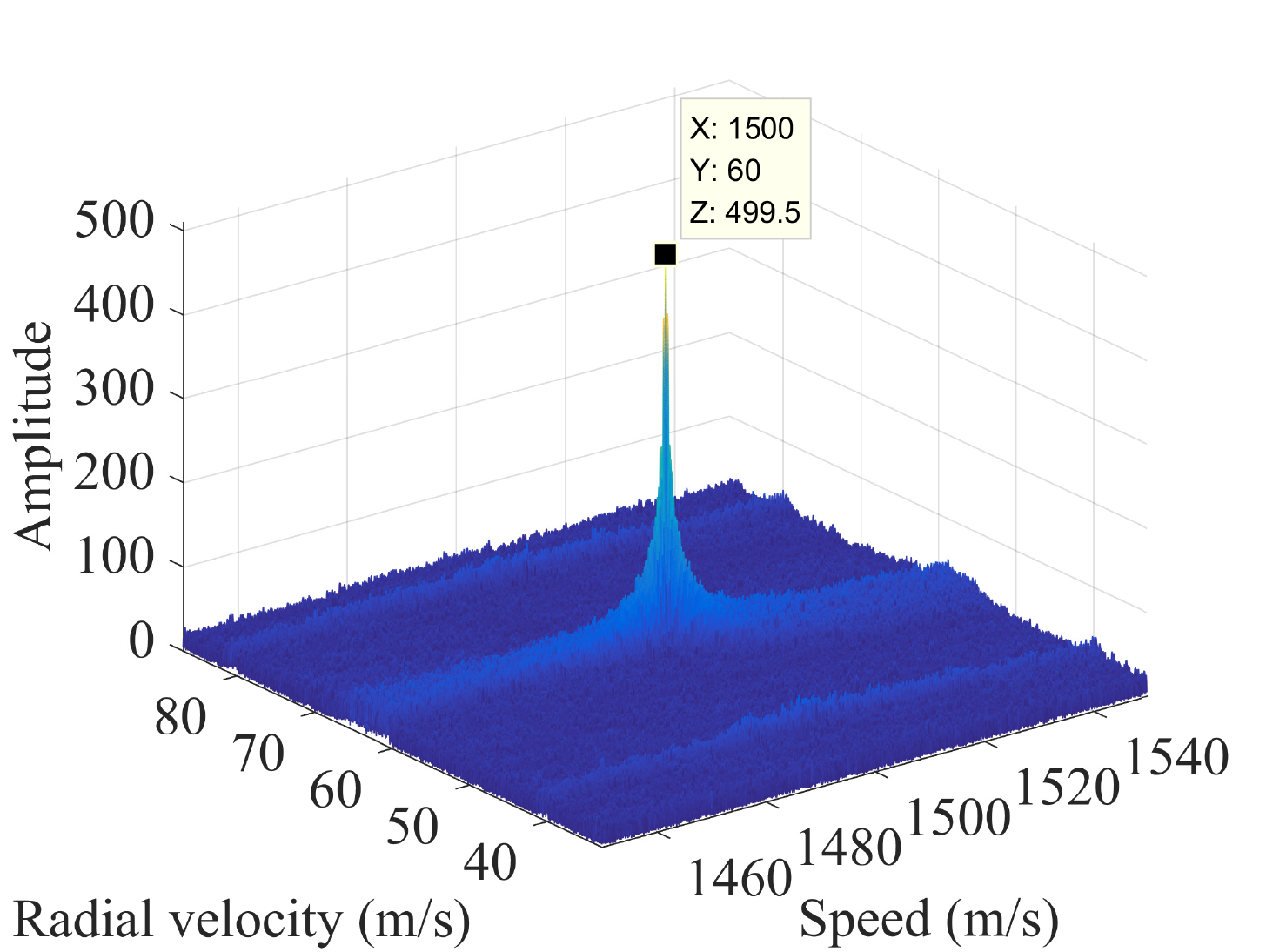} \label{R_slice4} }
	\subfigure[]{
		\includegraphics[width= 4.5 cm]{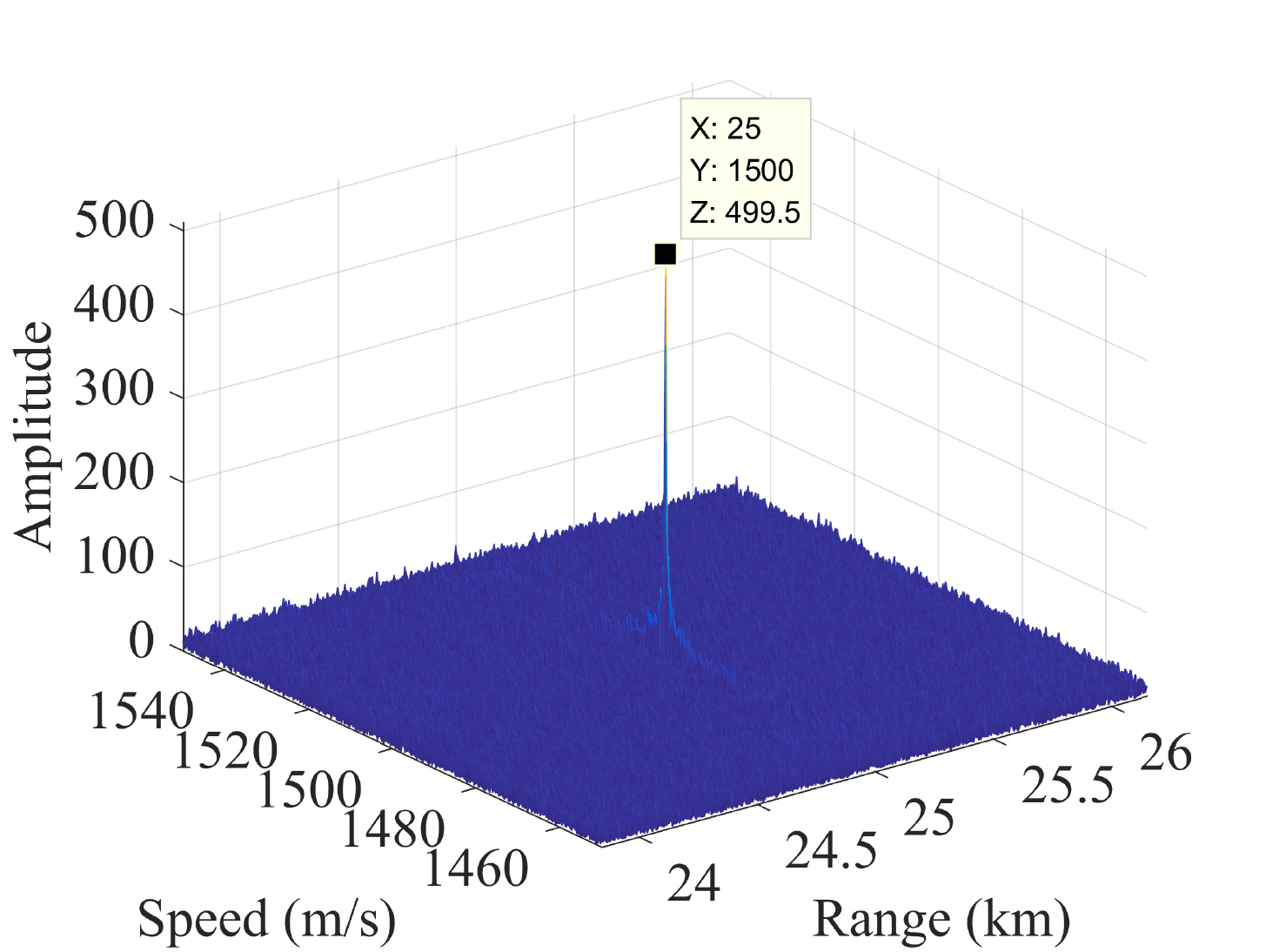} \label{V_slice4} }	
	\caption{Simulation results for high speed. 
		(a) Result after pulse compression. 
		(b) RFT. 
		(c) KT-MFP. 
		(d) GRFT. 
		(e) Range-Doppler slice of the AREM-GRFT at speed of 1500 $\mathrm{m} / \mathrm{s}$. 
		(f) Doppler-speed slice of the AREM-GRFT at range of 25 ${\rm{km}}$. 
		(g) Speed-range slice of the AREM-GRFT at radial velocity of 60 $\mathrm{m} / \mathrm{s}$.}
	\label{Simulation results for high speed.}
\end{figure}

\subsubsection{Performance in case of high speed}\label{AREM-GRFT of High Speed}
To evaluate the coherent integration performance of the proposed method for target with high speed, 
the target parameters are set as: 
initial range $r_{0}$ $=$ 25 $\mathrm{km}$, 
initial radial velocity $\dot{r}_{0}$ $=$ 60 $\mathrm{m} / \mathrm{s}$, 
speed $v$ $=$ 1500 $\mathrm{m} / \mathrm{s}$, 
and SNR after PC is 6 $\rm{dB}$. 
The radar parameters are the same as those in Table \ref{Table_Radar_Parameters}.

The corresponding simulation results in case of high speed are shown in Fig. \ref{Simulation results for high speed.}. 
In the case with the high speed, the compressed target signal also appears highly nonlinear with respect to time, 
as shown in Fig. \ref{PC4}. 
The RFT, KT-MFP, and GRFT suffer from serious performance loss 
and fail to produce effective integration of target energy, 
as illustrated in Fig. \ref{RFT4}--Fig. \ref{GRFT4}.

On the contrary, the proposed AREM-GRFT method provides effective integration output. 
A peak with amplitude of 499.5, close to the pulse number, 
is formed at the location corresponding to the true target parameters, 
as shown in Fig. \ref{VV_slice4}--Fig. \ref{V_slice4}. 
The proposed method obtains an almost ideal integration gain in this scenario.

\begin{table} [!t] \renewcommand{\arraystretch}{1.3} 
	\centering
	\caption{Parameters of the multiple targets.}
	\label{Table_Multi_Motion_Parameters}
	\setlength{\tabcolsep}{1.5mm}{
		\begin{tabular}{lcccc}
			\hline
			\hline
			Parameters (Unit)                				  	    &  Target 1     & Target 2      & Target 3     & Target 4    \\
			\hline
			Initial slant range (${\rm{km}}$)     				    &    21         &  21           &  21.15       &  21.21      \\
			
			Initial radial velocity ($\mathrm{m} / \mathrm{s}$)     &   -10         & -10           &  17          & -20         \\
			
			Speed ($\mathrm{m} / \mathrm{s}$)                       &   1150        &  1300         &  1300        &  1200       \\
			
			SNR after PC (${\rm{dB}}$)                              &   0           &  0            &  0           &  0          \\
			\hline
	\end{tabular}}
\end{table}

\subsection{Coherent integration for multiple targets}\label{Coherent Integration for Multiple Targets}

\begin{figure}[!t]
	\centering
	\subfigure[]{
		\includegraphics[width= 4.5 cm]{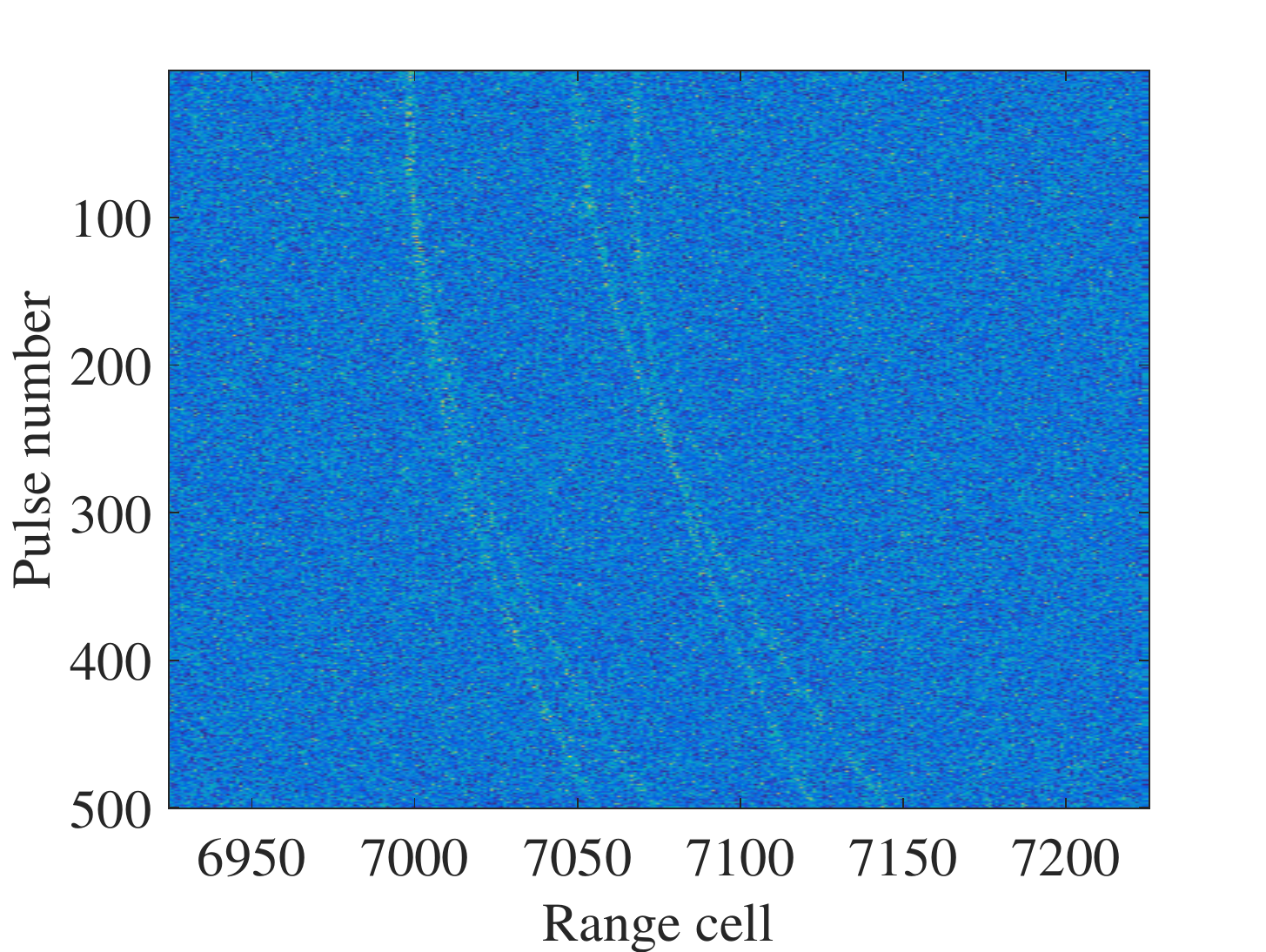} \label{PC52} }
	\subfigure[]{
		\includegraphics[width= 4.5 cm]{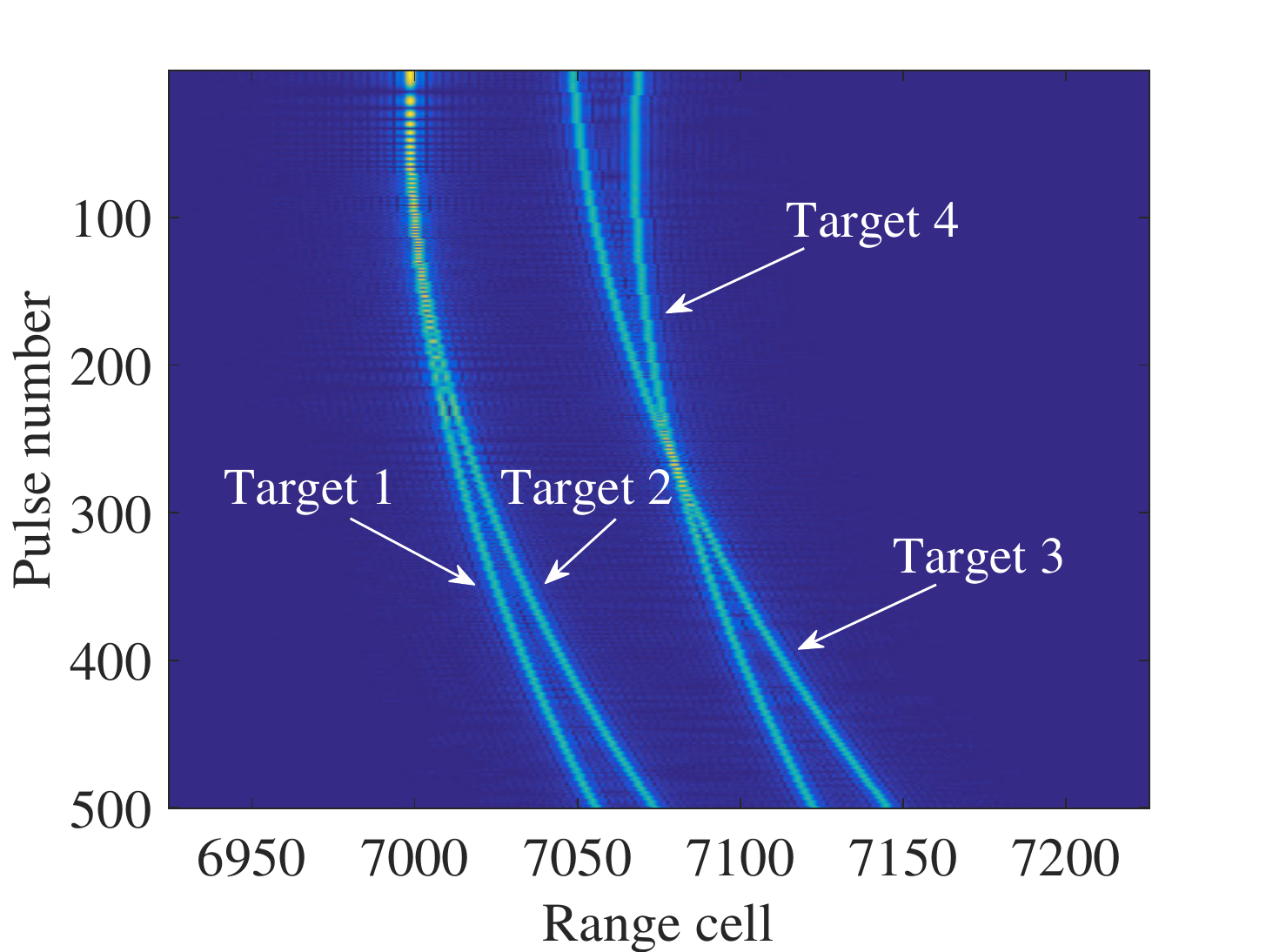} \label{PC5} }
	\subfigure[]{
		\includegraphics[width= 4.5 cm]{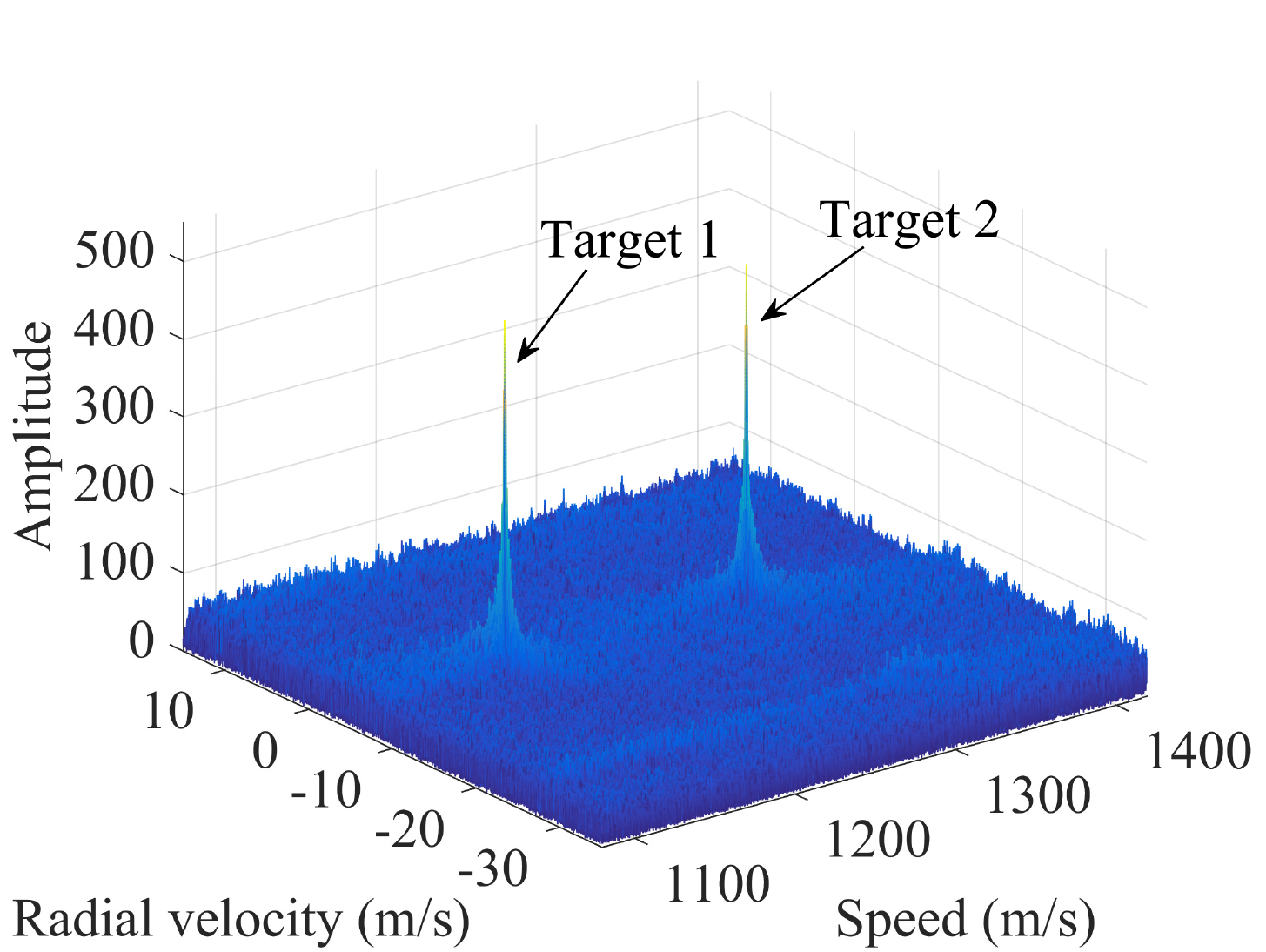} \label{R_slice5} }
	\subfigure[]{
		\includegraphics[width= 4.5 cm]{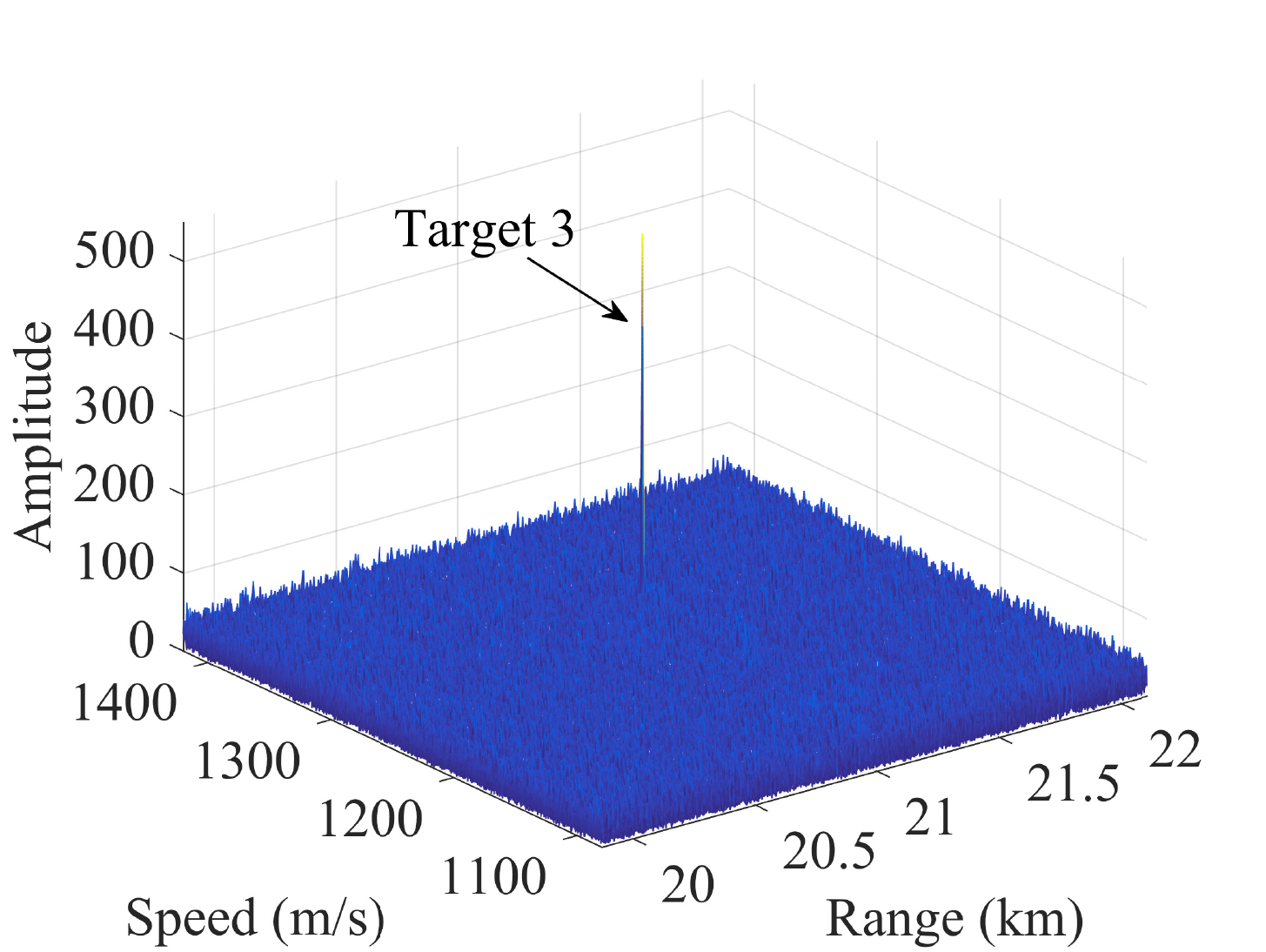} \label{V_slice5} }
	\subfigure[]{
		\includegraphics[width= 4.5 cm]{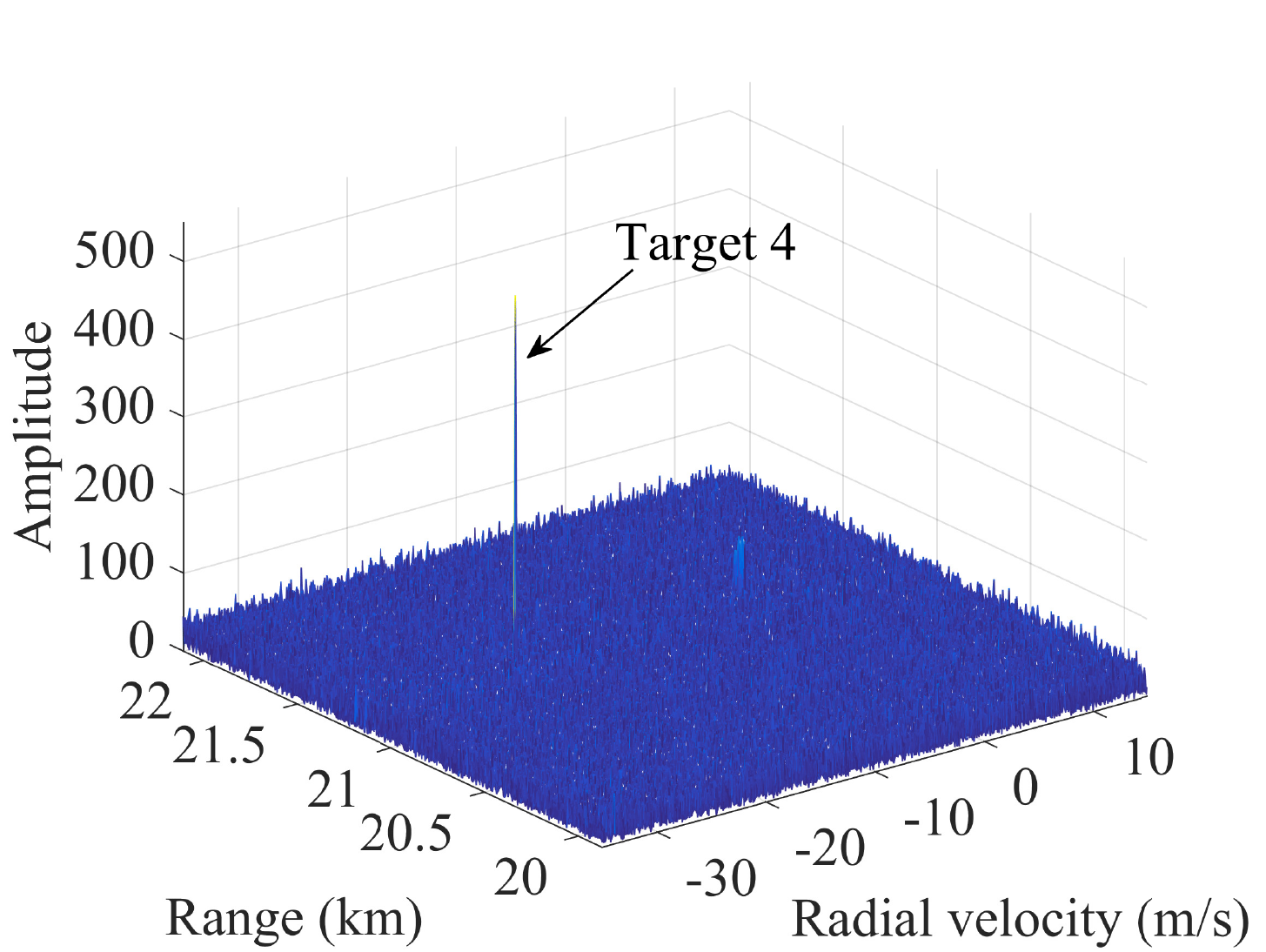} \label{VV_slice5} }
	\caption{Simulation result of the proposed method for multiple targets. 
		(a) Result after pulse compression. 
		(b) Result after pulse compression without noise. 
		(c) Doppler-speed slice of the AREM-GRFT at range of 21 ${\rm{km}}$. 
		(d) Speed-range slice of the AREM-GRFT at radial velocity of 17 $\mathrm{m} / \mathrm{s}$. 
		(e) Range-Doppler slice of the AREM-GRFT at speed of 1200 $\mathrm{m} / \mathrm{s}$.}
	\label{Simulation result of the proposed method for multiple targets.} 
\end{figure}

To evaluate the LTCI performance of the proposed method for multiple targets, 
four targets with CCV motion are considered in this scenario, 
whose parameters are given in Table \ref{Table_Multi_Motion_Parameters}. 
Target 1 and target 2 have the same initial range and radial velocity, but different speeds. 
The initial range and radial velocity of target 3 are different from those of target 2, 
and target 4 has parameters different from the parameters of all the other targets. 
The radar parameters are the same as those in Table \ref{Table_Radar_Parameters}.

Fig. \ref{Simulation result of the proposed method for multiple targets.} shows the simulation results of the AREM-GRFT method for multiple targets. 
The signal after pulse compression is shown in Fig. \ref{PC52}. 
It can be seen that the signal energy is almost submerged in noise. 
In order to clearly show the trajectories of the targets, 
the pulse compression result without noise is shown in Fig. \ref{PC5}. 
The slice of Doppler-speed at range of 21 ${\rm{km}}$, 
identical to the initial range of target 1 and target 2, 
is illustrated in Fig. \ref{R_slice5}. 
Two peaks can be observed, 
and the peak positions are consistent with target 1 and target 2, respectively. 
Note that these two peaks have the same radial velocity. 
They can only be distinguished in the direction of speed. 
In Fig. \ref{V_slice5}, the slice of speed-range with radial velocity 17 $\mathrm{m} / \mathrm{s}$, 
identical to Target 3 is provided. 
In Fig. \ref{VV_slice5}, 
the slice of range-Doppler at speed of 1200 $\mathrm{m} / \mathrm{s}$ is illustrated. 
In each of these two figures, only one peak is observed. 
This is because only one target has the same parameter corresponding to the slice. 
The peak in Fig. \ref{V_slice5} is target 3 and that in Fig. \ref{VV_slice5} is target 4. 
Based on the integration results in Fig. \ref{Simulation result of the proposed method for multiple targets.}, 
it can be concluded that the proposed AREM-GRFT method could not only achieve effective coherent integration for multiple targets, 
but also provide additional observation and resolution in speed domain.

It should be noted that like some parameter-search algorithms such as RFT \cite{RFT1-2011} and GRFT \cite{GRFT-2012}, 
the blind speed sidelobe (BSSL) also exists in the proposed method 
(see Fig. \ref{VVslice_radial_motion}, Fig. \ref{VV_slice2}, Fig. \ref{VV_slice4} and Fig. \ref{VV_slice5}). 
When the searching radial velocity differs from the radial velocity of the target by an integer times of blind speed, 
the target energy would also be partially integrated. 
Some effective methods \cite{RFT2-2011, IMWRFT-2020} have been proposed for BSSL suppression of RFT. 
To suppress the BSSL in the proposed method is an important topic needs to be further studied.

\subsection{Detection performance analysis}\label{Detection Performance Analysis}
In this subsection, the detection performance of the proposed AREM-GRFT method and 
several typical coherent integration methods (i.e., MTD, RFT, KT-MFP, GRFT) are investigated by Monte Carlo experiments. 
The radar parameters and motion parameters of the target are the same as those in Section \ref{AREM-GRFT of Long Integration Time}. 
The CFAR detector is used for the five coherent integration methods, 
i.e., MTD, RFT, KT-MFP, GRFT, and the proposed AREM-GRFT method. 
The false alarm probability is set as $P_{f a}=10^{-4}$ and the Gaussian noises are added to the target echoes. 
The SNR after pulse compression varies from $-$40 ${\rm{dB}}$ to 30 ${\rm{dB}}$ and the step size is 1 ${\rm{dB}}$. 
In each case, 1000 times Monte Carlo experiments are carried out.

\begin{figure}[!t]  
	\centering  
	\includegraphics[scale= 0.55 ]{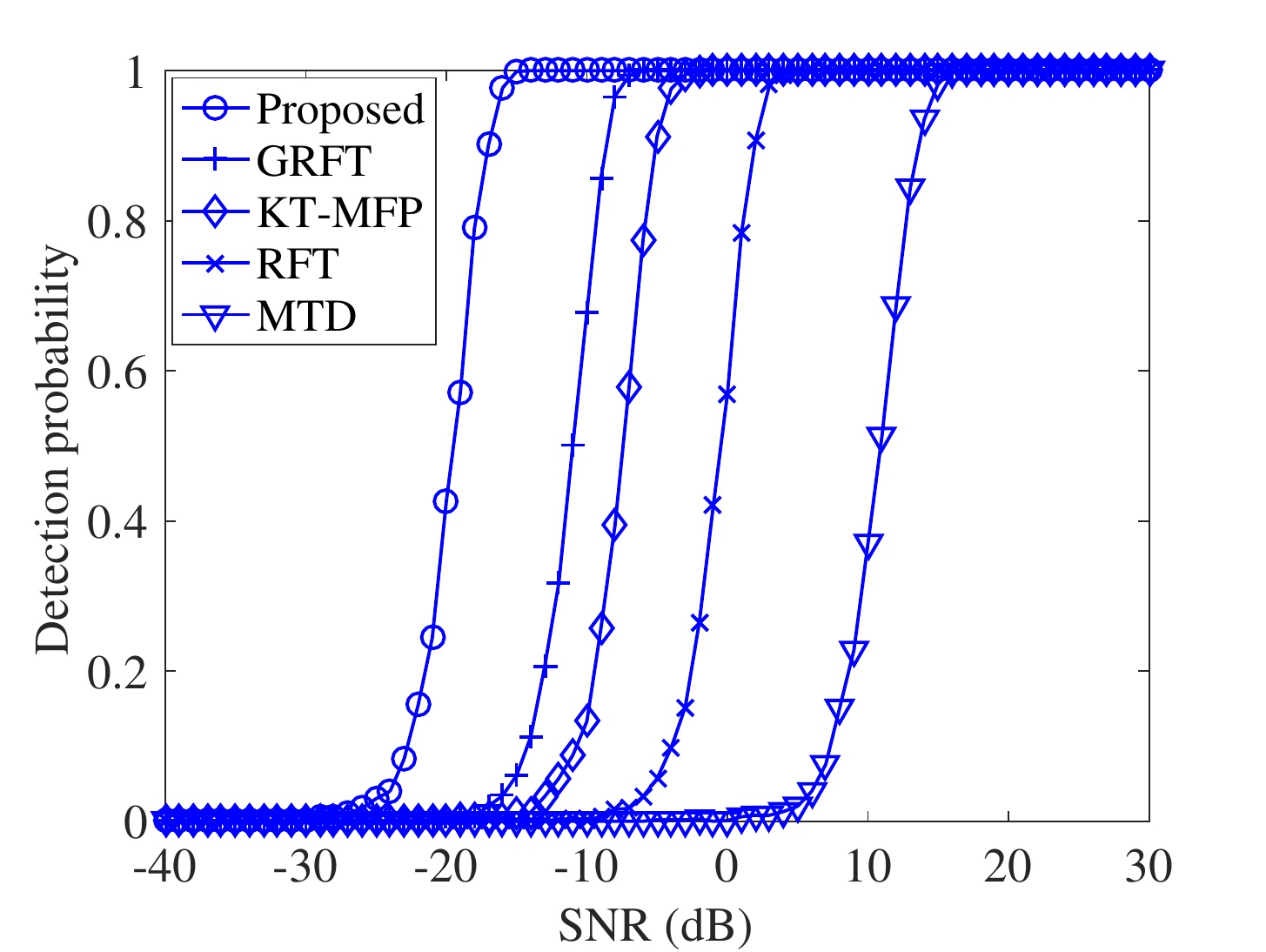}  
	\caption{Detection probability of MTD, RFT, KT-MFP, GRFT, and the proposed method.}   
	\label{Detection_probability}
\end{figure}

The detection probability under different SNR levels is shown in Fig. \ref{Detection_probability}. 
It can be seen that the detection performance of the proposed method is superior to MTD, RFT, KT-MFP, and GRFT. 
Particularly, for the detection probability ${P_d}$ $=$ 0.8, 
the required SNR of AREM-GRFT is about 8$/$13$/$20 ${\rm{dB}}$ lower than that of GRFT$/$KT-MFP$/$RFT, respectively. 
It is because that the proposed method can achieve effective energy integration 
thanks to the ability to match CCV motion accurately, 
while MTD, RFT, KT-MFP, and GRFT fail to deal with the highly nonlinear RM and complex DFM 
due to model mismatch and inaccurate energy accumulation.

%%%%%%%%%%%%%%%%%%%%%%%%%%%%%%%%%%%%%%%%%%%%%%%%%%%%%%%%%%%%%%%%%%%%%%%%%%%%%%%%Conclusion
\section{Conclusion}\label{Conclusion}
In this paper, 
a novel generalized Radon Fourier transform method for target moving with constant Cartesian velocities (CCV) was presented, 
based on the accurate range evolving model, 
which is a square root of a polynomial with terms up to second order with target speed as the factor. 
The accurate model instead of approximate polynomial models used in the proposed method enables 
effective energy integration on characteristic invariant with feasible computational complexity. 
The target samplings are collected and the phase fluctuation among different pulses is compensated 
according to the accurate range evolving model. 
The highly nonlinear range migration and complex Doppler frequency migration caused by CCV motion are eliminated simultaneously. 
Numerical simulation experiments demonstrated that the proposed method can 
achieve better integration gain and detection performance than 
several typical polynomial model based coherent integration methods (i.e., RFT, KT-MFP and GRFT). 
Additionally, the proposed method can not only achieve effective coherent integration for CCV targets 
regardless of the coverage of target speed and coherent processing interval, 
but also provide additional observation and resolution in speed domain.

% \section*{Acknowledgment}

%\section{References}
%\bibliographystyle{ieeetr}
%\bibliography{totalReference}

%\newpage

\end{document}